\documentclass[useAMS,usenatbib]{mn2e_mod}
\usepackage{graphicx}
\usepackage{amsmath}
\usepackage{amssymb}
\usepackage{deluxetable}
\usepackage{longtable}

\title [The weirdest SDSS galaxies]{The weirdest SDSS galaxies: results from an outlier detection algorithm}

\author[Baron et al.]
{Dalya Baron$^{1}$\thanks{dalyabaron@mail.tau.ac.il},
Dovi Poznanski$^{1}$\thanks{dovi@tau.ac.il}\\
$^{1}$School of Physics and Astronomy, Tel-Aviv University, Tel Aviv 69978, Israel.}

\begin{document}

\maketitle

\label{firstpage}
\begin{abstract}

How can we discover objects we did not know existed within the large datasets that now abound in astronomy? We present an outlier detection algorithm that we developed, based on an unsupervised Random Forest. We test the algorithm on more than two million galaxy spectra from the Sloan Digital Sky Survey and examine the 400 galaxies with the highest outlier score. We find objects which have extreme emission line ratios and abnormally strong absorption lines, objects with unusual continua, including extremely reddened galaxies. We find galaxy-galaxy gravitational lenses, double-peaked emission line galaxies, and close galaxy pairs. We find galaxies with high ionisation lines, galaxies which host supernovae, and galaxies with unusual gas kinematics. Only a fraction of the outliers we find were reported by previous studies that used specific and tailored algorithms to find a single class of unusual objects. Our algorithm is general and detects all of these classes, and many more, regardless of what makes them peculiar. It can be executed on imaging, time-series, and other spectroscopic data, operates well with thousands of features, is not sensitive to missing values, and is easily parallelisable.

\end{abstract}

\begin{keywords}
galaxies: general -- galaxies: peculiar -- methods: machine learning -- methods: data analysis -– methods: statistical

\end{keywords}

\vspace{1cm}
\section{Introduction}\label{s:intro}

The amount of astronomical data and its complexity grows rapidly, thus introducing astronomy to the era of Big Data science. Ongoing and future large surveys provide astronomers with a wealth of information to be analysed and extracted from rich and complex data types. The Sloan Digital Sky Survey (SDSS) has provided the community with multi-coloured images of approximately one third of the sky, and spectra of more than three million objects \citep{york00}. The Palomar Transient Factory (PTF; \citealt{law09}) and Pan-STARRS \citep{kaiser10} monitored the variable night sky and studied numerous asteroids, variable stars, supernovae, active galactic nuclei (AGN), and more. Future surveys will increase by orders of magnitude the number of available objects and the information each of these holds. This includes most notably LSST \citep{ivezic08}, DESI \citep{levi13}, and ZTF \citep{bellm14}. These challenge astronomers to move on from the traditional data analysis and visualisation techniques and develop new, Big-Data appropriate, tools \citep{pesenson10}.

In view of this accelerating growth, astronomers start to develop automated tools in order to detect, characterise, and classify objects based on the immense amount of information they are gathering. Following that, it is necessary to point towards objects which hold the largest amount of new information by developing automated decision making tools that can process large numbers of objects and draw our specific attention. Supervised Machine Learning (ML) algorithms are used to detect predefined objects of interest, and unsupervised learning algorithms to classify objects and find correlations in high dimensional data (\citealt{collister04, fiorentin07, daniel11, luis11, richards12, almeida13, fustes13, masci14, baron15, hocking15, miller15, wright15, djorgovski16, lochner16, mackenzie16, rubin16}; for additional references and review see \citealt{ball10}).

The natural step after classification and the analysis of different clusters is outlier detection and characterisation. Outliers can be objects that were not intended to be included in the original sample, various errors, extreme objects that reside on the tail of well characterised distributions, and completely new objects which offer the opportunity to unveil new physics. To paraphrase Donald Rumsfeld\footnote{https://www.youtube.com/watch?v=GiPe1OiKQuk}: normal objects are the `known knowns'; outliers of kinds discovered before are `known unknowns'; but the most interesting, and challenging to find in large datasets, are the `unknown unknowns', objects we did not know we should be looking for. Our goal in this paper is to present such an outlier detection algorithm and its application on the well studied sample of SDSS galaxy spectra, as a first test.

Random Forest (RF) \citep{breiman84, breiman01} is being used extensively in astronomy, though as a supervised learning algorithm. Its main advantages are that it offers a feature importance ranking, and it can be easily parallelisable. Studies train RF to distinguish between different classes of objects based on their features. This is being done by construction of a sample in which objects are already (manually or via a different method) labeled. This sample is the training sample with which the algorithm is trained to distinguish between different classes of objects. These decision rules are then applied to a sample of interest with no labels. RF is used for classification of transients in PTF \citep{bloom12} and in the Dark Energy Survey \citep{goldstein15}, and it was recently proposed for gravitational wave searches of black hole binary coalescence \citep{baker15}. 

Our outlier detection is based on an unsupervised RF, thus we do not have labeled data, and expect the algorithm to learn the data through its features with no additional user-defined input. Our algorithm is based on the work of \citet{shi06}, but with modifications that make it suitable for this task. For other anomaly detection algorithms see \citet[e.g., the work of \citealt{liu08}]{goix16}. We test the algorithm on one of the most extensively studied samples - the SDSS galaxy spectra. The twelfth data release (DR12) of SDSS includes spectra of more than 2 million galaxies, approximately half of which were released to the public 7 years ago. These galaxies were characterised and classified thoroughly, and unusual objects were found, making it a useful benchmark (see for example: \citealt{almeida10, andrae10, ascasibar11, huertas11}). 

\begin{figure*}
\includegraphics[width=0.8\textwidth]{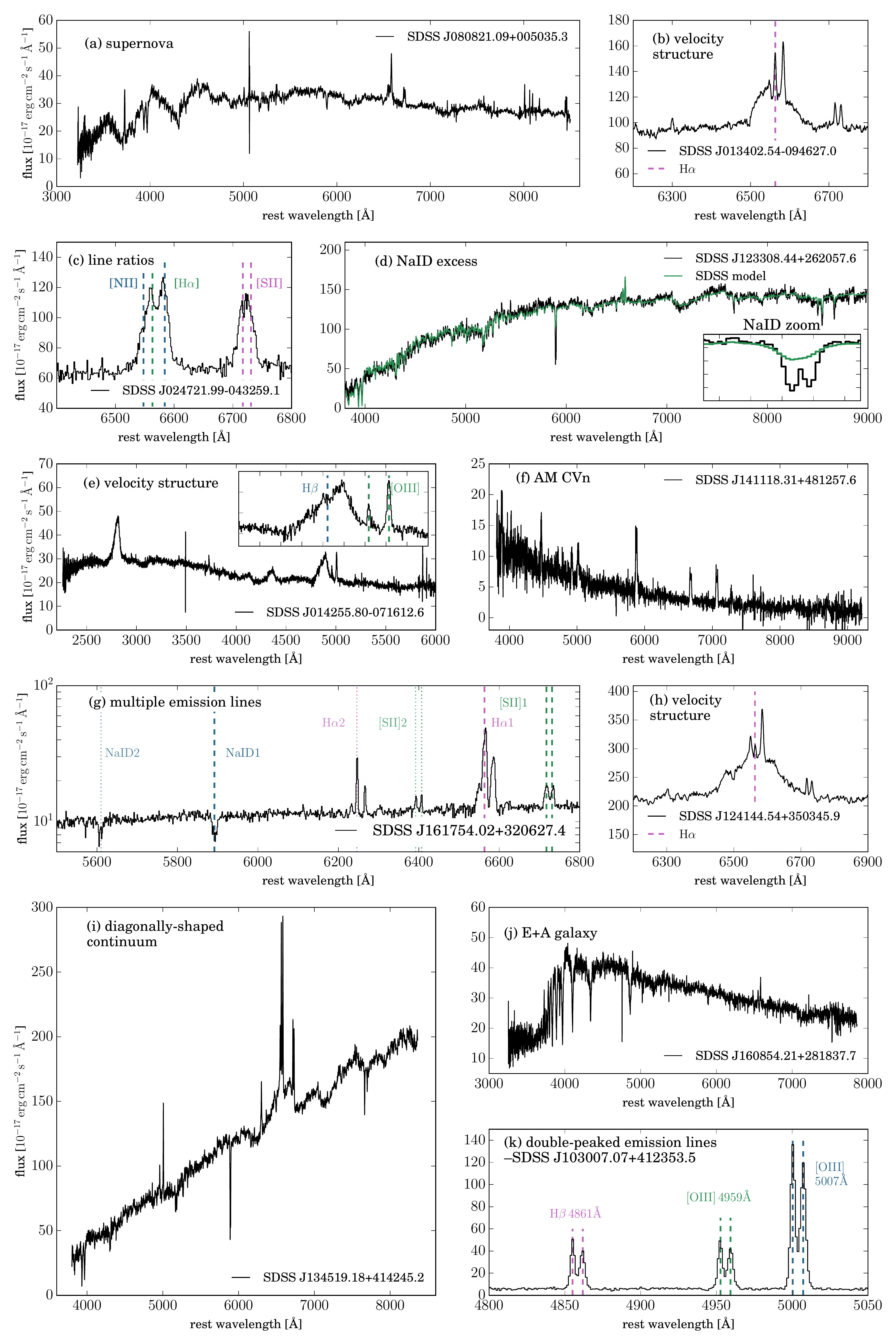}
\caption{
A sample of the outliers we find with the outlier detection algorithm, sporting a large variety of unusual features, all found within a single run of the algorithm.
}\label{f:uber_plot}
\end{figure*}

As we show below, we find different kinds of outliers which were found by previous studies, but more importantly, some of the outliers we find are new, even in a sample as thoroughly studied as this one. We show in figure \ref{f:uber_plot} a sample of outlying galaxies that our algorithm detects, some of them are further discussed in later sections. Clearly, we find objects with a wide variety of properties. 

We describe the galaxy sample in section \ref{s:data} and present the outlier detection algorithm in section \ref{s:algo}. We then present and discuss the 400 weirdest spectra found by the algorithm in section \ref{s:outlier}. We conclude with a discussion regarding other potential uses of the algorithm in astronomy in section \ref{s:conc}.

\section{Data Set}\label{s:data}
Our sample consists of all the galaxy spectra from SDSS DR12 \citep{alam15}, obtained with both the SDSS and BOSS spectrographs (see \citealt{ahn12}). Although the spectrographs differ in fibre diameter and wavelength coverage we do not divide our sample based on the spectrograph that was used. 
We have 2\,379\,168 galaxy spectra in total. 

In order to avoid being confused by bad sky subtractions and calibration errors, we use the `and{\_}mask' which marks pixels that are bad throughout all the sub-exposures of a given object. We discard pixels that are marked in `and{\_}mask' and their closest neighbours (one on each side). We completely remove all objects which have more than 700 bad pixels marked in the `and{\_}mask' and remain with 2\,355\,926 spectra. We further discard the pixels in the wavelength range 5565-5590 \AA\, in the observers frame due to residual sky emission which occurs in a large number of the objects.

The SDSS spectroscopic pipeline obtains the redshift of an object by cross-matching its spectrum to template spectra of stars, galaxies, and quasars. However, in BOSS, galaxies are often classified as quasars with unrealistic fit parameters and the original redshift estimation is incorrect. Therefore, another fit is done, with no quasar templates, which yields a better fit and redshift estimation, $z_{\mathrm{noqso}}$ \citep{bolton12}. For every object with $z_{\mathrm{noqso}}$ available, we use it instead of the standard redshift. We shift the spectra to the rest-frame wavelength, and then interpolate the spectra to an identical wavelength grid from 2250 to 10100\AA\, with 0.5\AA\, resolution, and normalise the flux using the median flux of the spectrum. Redshift is a very nonlinear variable, and working in the observer frame would complicate any classification task significantly. Here domain knowledge allows us to remove the effect of redshift in pre-processing. It is important to note that, as we show below, objects with erroneous redshifts assigned by the SDSS pipeline are found naturally as outliers by our algorithm. We do not extrapolate, and use NULL values at wavelength where we do not have information.  We finally apply a running median filter with a window size of 2.5\AA\, (i.e., 5 pixels) to pixels that are not NULL, which is roughly the resolution of the spectrograph. This is done in order to remove residual cosmic-rays and noise. The width of the filter is small enough to have little effect on real spectral lines.
Overall, we perform a limited pre-analysis of the data and the steps detailed above only discard 1\% of the data.

\begin{figure*}
\includegraphics[width=3.25in]{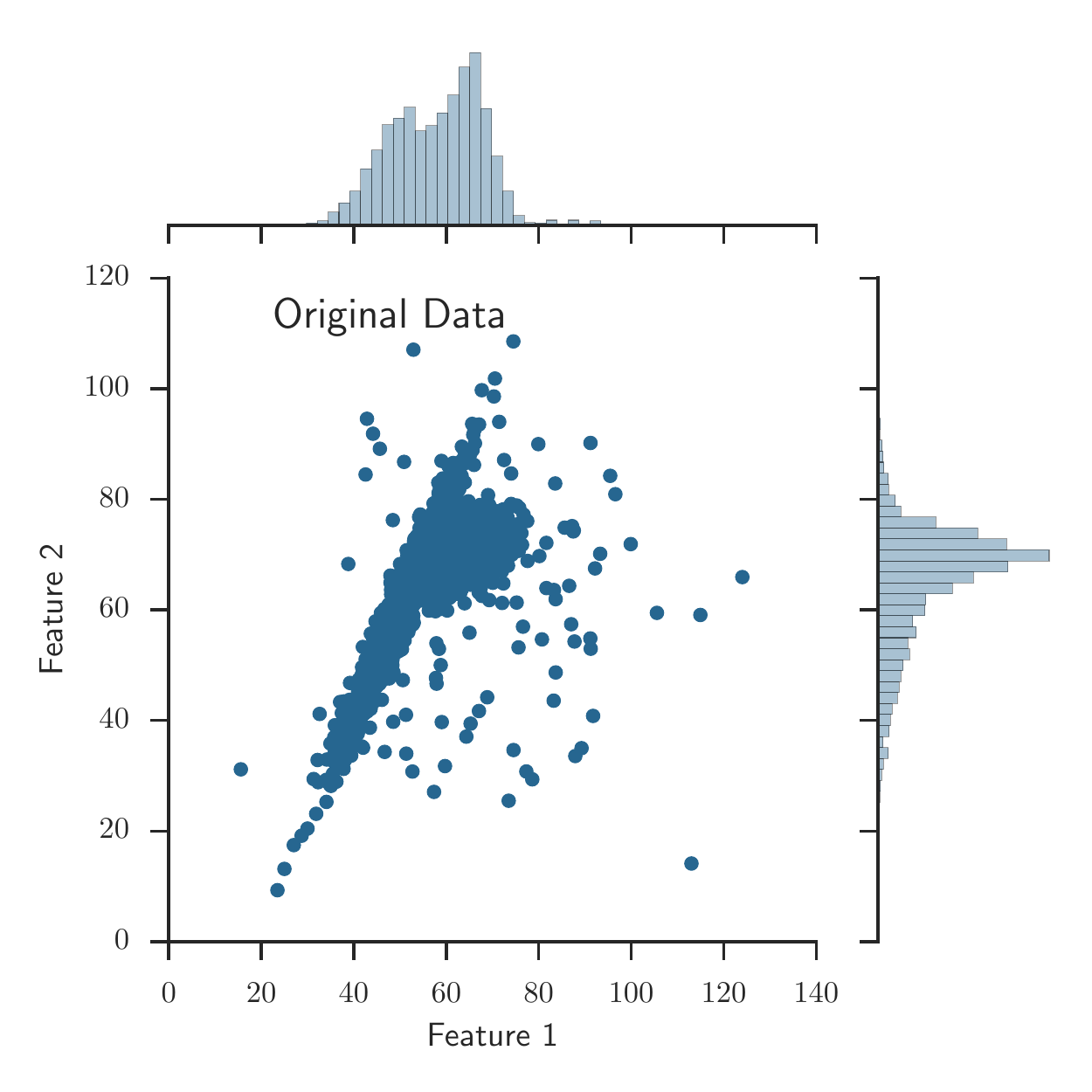}
\includegraphics[width=3.25in]{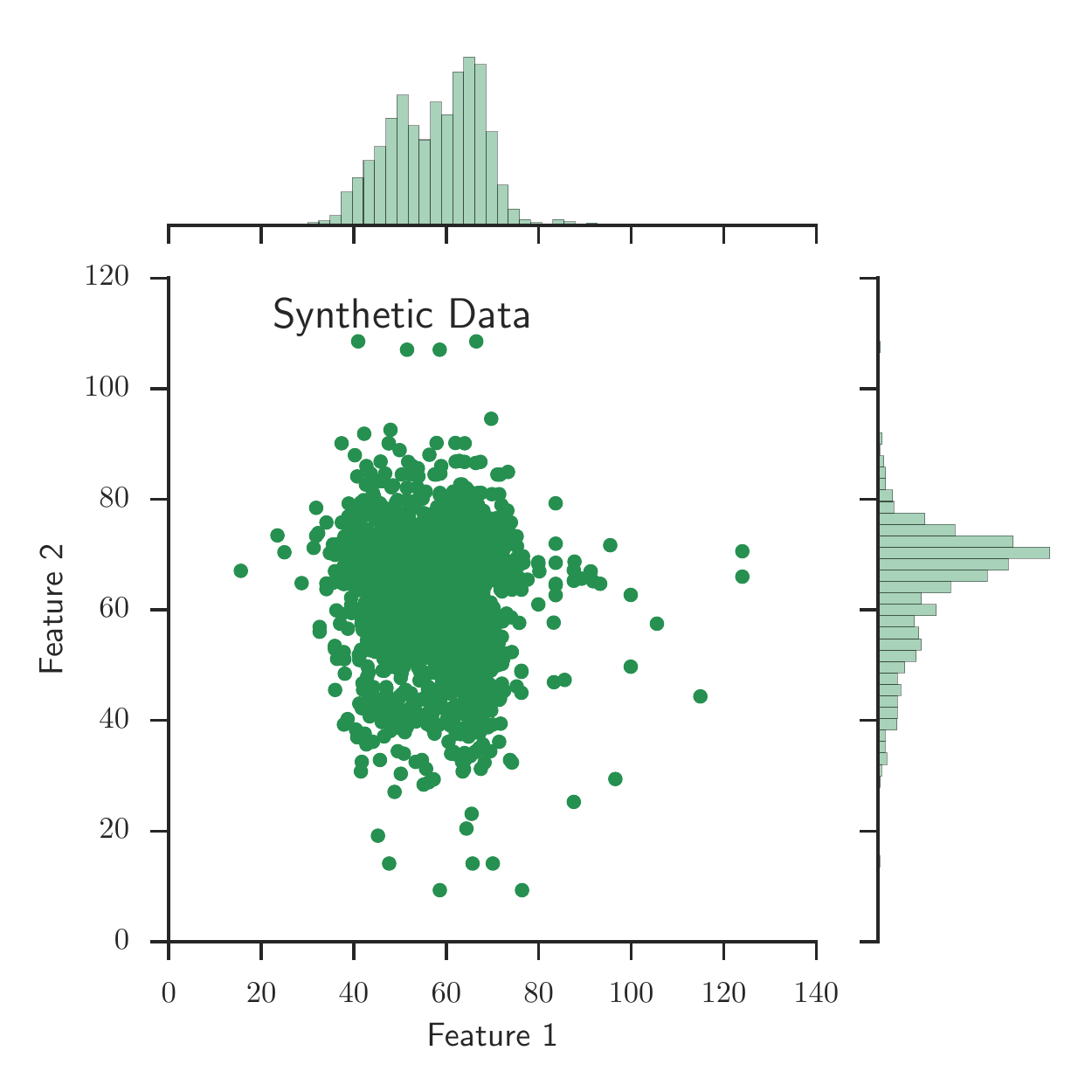}
\caption{
Illustration of the synthetic data construction: we sample points from the marginal distributions of the original data (left panel) and define it as the synthetic data (right panel) for a given feature (which is a single flux measurement). While the marginal distribution of the two samples remain identical, the covariance (i.e., the dependence between the different features) exists only in the original data.
}\label{f:syn_data_illus}
\end{figure*}

\section{Outlier Detection Algorithm}\label{s:algo}
The algorithm we discuss is general and can be implemented on images, time-series (such as light curves), and photometry, similarly to our current application to spectroscopic data. The input to the algorithm is a matrix in which the rows represent the objects in the sample (galaxy spectra in our case) and the columns represent the features of the data. In our case, the features are simply the flux values at each wavelength, thus we have 2\,355\,926 objects with 15\,700 features each (i.e., 15\,700 interpolated flux values).

\begin{figure*}
\includegraphics[width=1\textwidth]{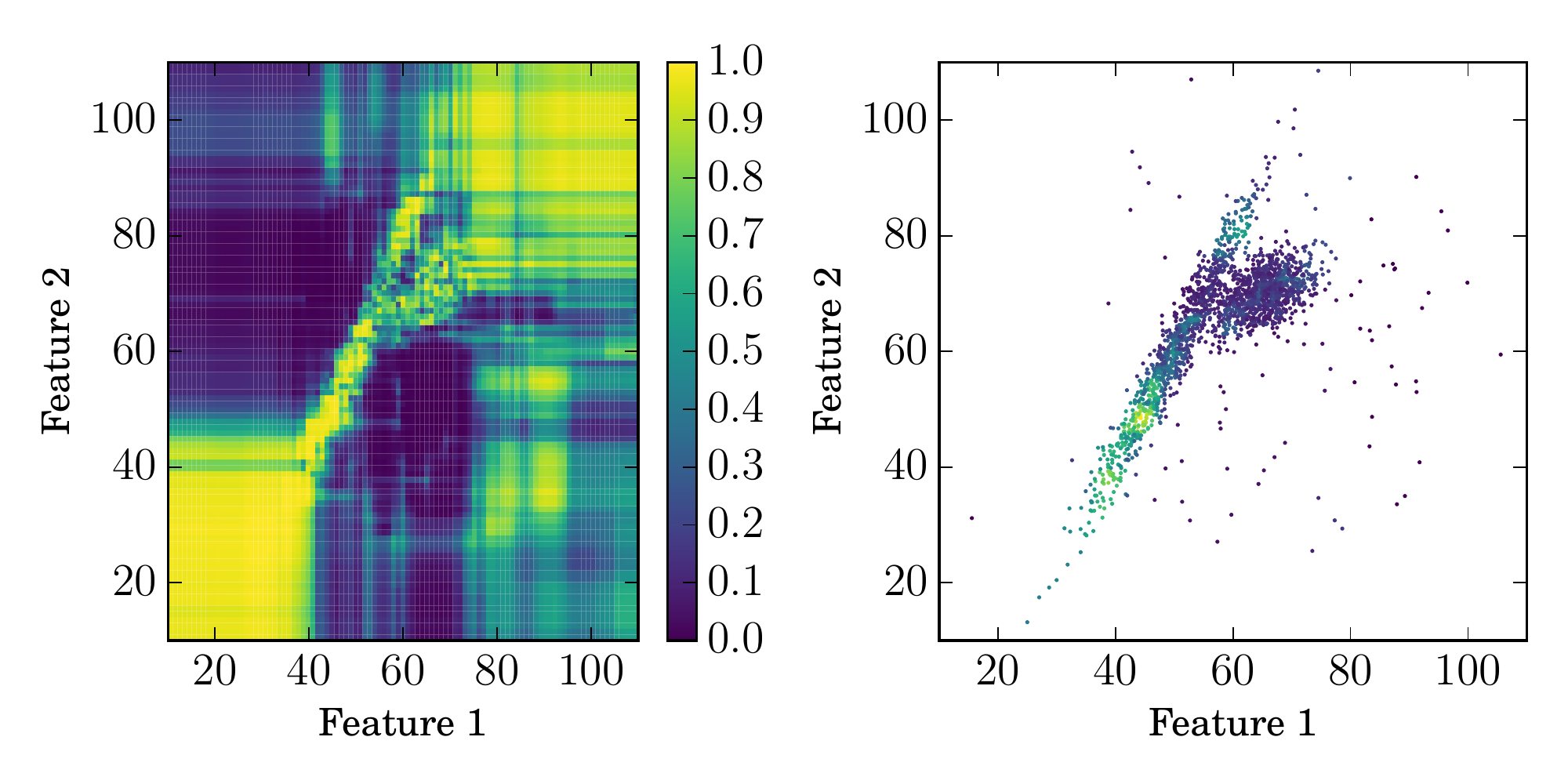}
\caption{Application of the algorithm to the illustration data from figure \ref{f:syn_data_illus}. The left panel shows a colormap of the probability of an object to be classified as \emph{real} by the trained RF. One can see that although the \emph{real} and \emph{synthetic} data occupy the same region in feature space, the RF provides good predictions for the location of \emph{real} objects. The right panel shows the \emph{real} objects, their colour represents their weirdness score, where dark blue represents high weirdness score and yellow represents a low weirdness score.
}\label{f:weirdness_score_color}
\end{figure*}

In order to detect outliers one can define a distance measure between every two objects in the sample, we achieve this task by using RF. An RF refers to the general technique of random decision forests, an ensemble learning method that is used for regression and classification. RF operates by constructing a set of decision trees during training. Every such tree splits the training data into the known classes by using rules learned from the data features. Each tree in the ensemble is built from a sample drawn with replacement from the objects in the sample and only a randomly selected subset of the features is used as a splitting criterion. Then, the RF averages the predictions of the decision trees, and due to this randomness the resulting model is general and has low variance \citep{breiman84, breiman01}. Constructing each decision tree with a subset of the objects and a subset of the features makes the RF naturally parallelisable and extremely fit for usage in the framework of Distributed File Systems (DFS).

Since the objects in our sample are not labeled and there is no external information input to the algorithm from the user, we perform unsupervised learning with RF, as described in \citet{shi06} though with some key modifications. We construct a synthetic data matrix with the same dimensions as our sample data (2\,355\,926 X 15\,700): each feature (column) in the synthetic data is built by sampling from the marginal distribution of the same feature in the original data. The process of creating the synthetic data is illustrated in figure \ref{f:syn_data_illus} in a simplified example, where each object has only two features, distributed as seen in the left panel. The synthetic data is shown in the right panel, where one can see that the marginal distributions of the original and synthetic data match, while the covariance between the features remains only in the original data.

Once we have the synthetic data, we can translate the problem to the language of supervised learning -- we label the original data as \emph{real} and the synthetic data as \emph{synthetic}. The total labeled data which consists of both the \emph{real} and \emph{synthetic} samples is the input to the traditional RF classification algorithm. We note that there is no need to divide the sample into training, validation, and test sets as typically done when performing supervised learning, since we do not use the trained forest as a predictor for new, unlabelled, measurements. We train the forest to distinguish between \emph{real} and \emph{synthetic}, and since the difference between the classes is the lack of covariance in the \emph{synthetic} data, the forest identifies features which show strong covariance and rates them higher in its features importance scheme. In this framework, spectral lines will show as correlations on short scales, between nearby wavelengths, while continuum slopes manifest themselves as more subtle correlations (typically) between distant wavelength ranges. Noise is largely uncorrelated, except perhaps for the influence of the point-spread-function that introduces short-range correlations in spectra. We apply the RF training to the illustration data from figure \ref{f:syn_data_illus} and show in the left panel of figure \ref{f:weirdness_score_color} the probability of an object to be classified as \emph{real} by the trained RF. One can see that although the \emph{real} and \emph{synthetic} data occupy the same region in parameter space, the RF learns the region of the \emph{real} objects well.

\begin{figure}
\includegraphics[width=3.25in]{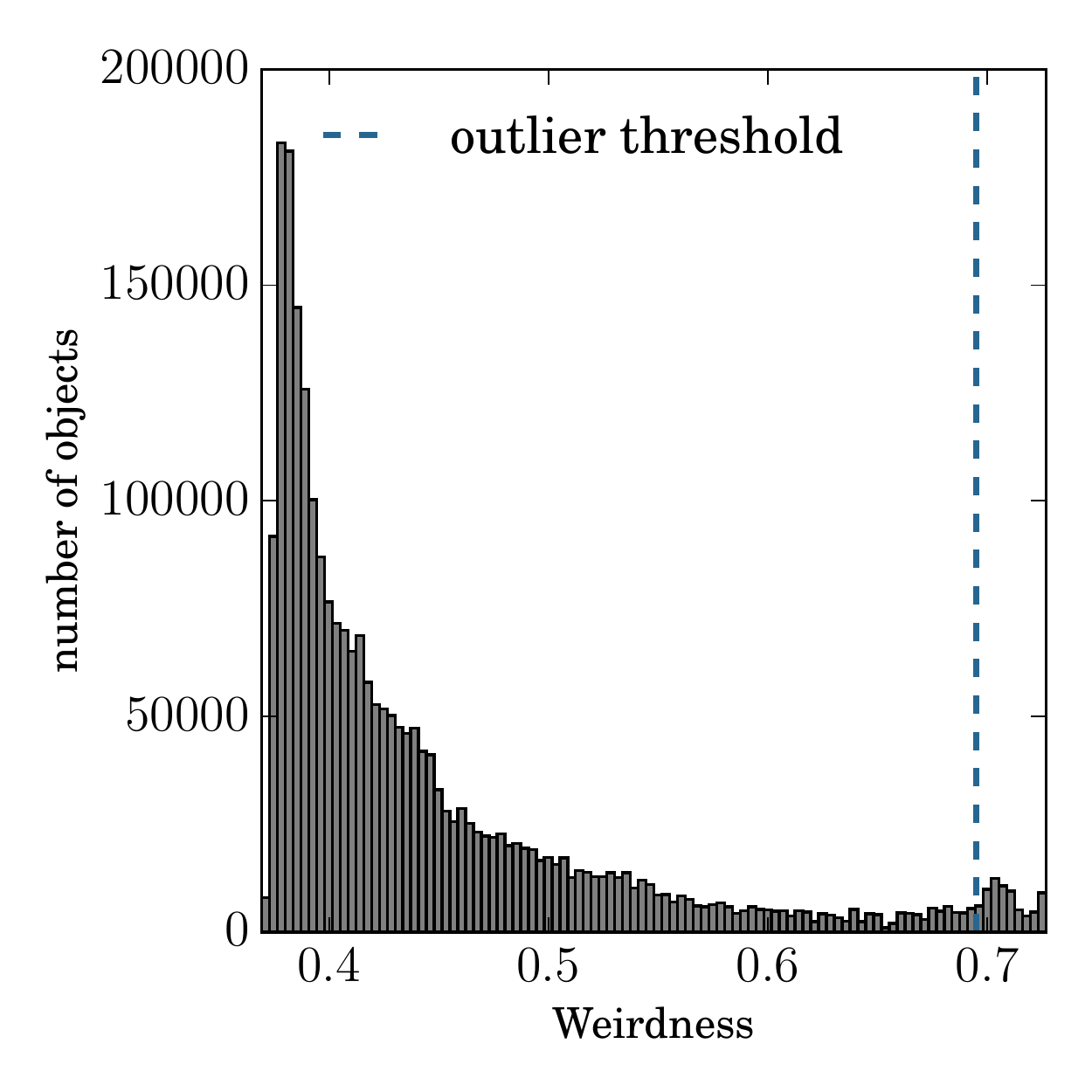}
\caption{
Histogram of the weirdness score for every object in our sample. One can see that the distribution is smooth and decreases as one approaches large values. We study in this paper the 400 objects that lie to the right of the dashed blue line.
}\label{f:dist_hist}
\end{figure}

\begin{figure}
\includegraphics[width=3.25in]{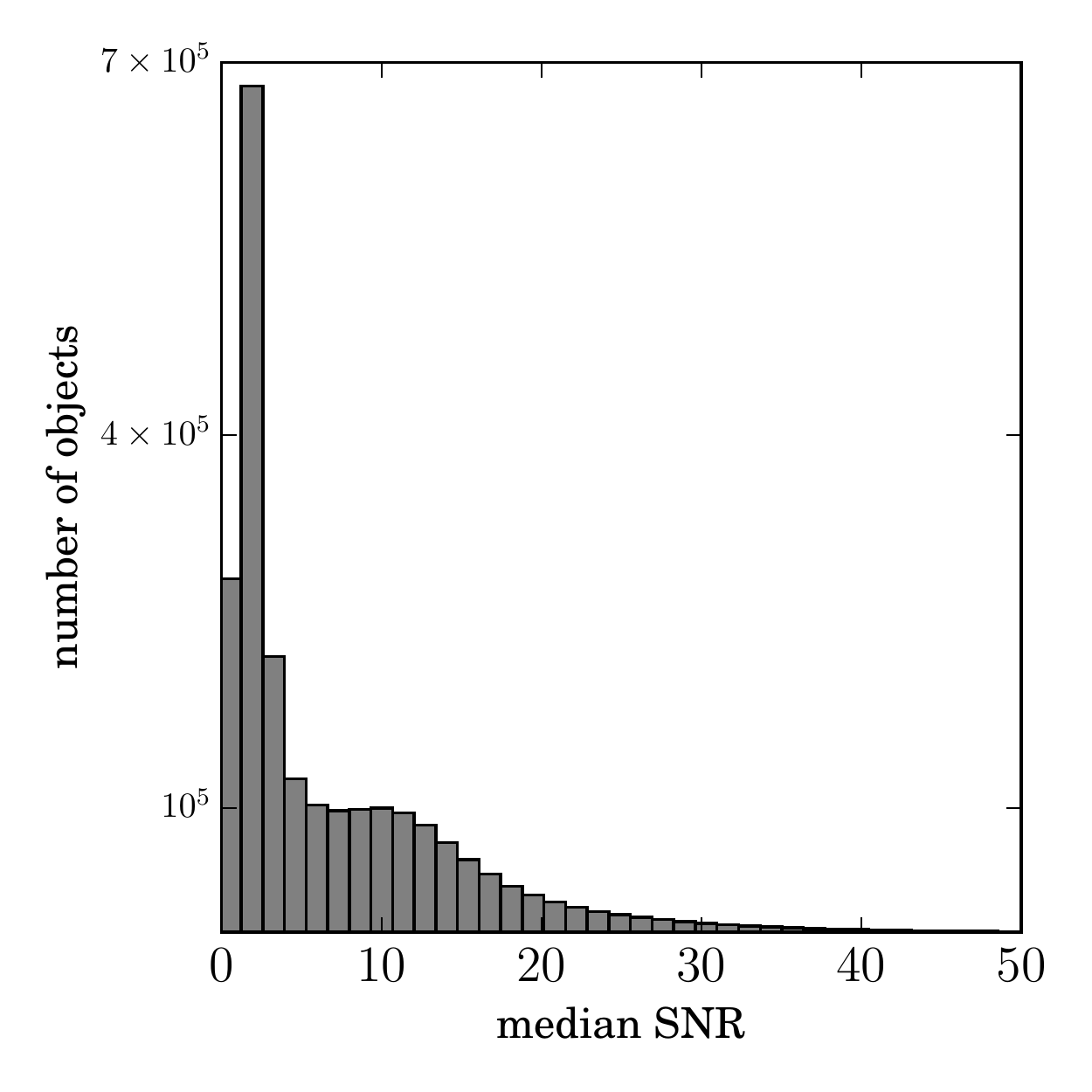}
\caption{
Histogram of the signal to noise ratio (SNR) of the galaxies in our sample. We define the SNR as the median SNR per pixel, as given by the SDSS pipeline. One can see that apart from a prominent peak in the distribution at SNR$\sim$2.5, the number of galaxies decreases smoothly for increasing SNR.
}\label{f:snr_distribution}
\end{figure}

Having the trained forest, we take only the \emph{real} objects and propagate them through the decision trees in the forest. Every object, in every tree, ends up in a terminal node (leaf) with a predicted class, either \emph{real} or \emph{synthetic}. In order to define a measure of similarity between objects, we count how often every pair of objects were classified as \emph{real} in the same leaf of a given tree. This number, $\mathrm{N_{leaf}}$, can range from 0 to the number of trees in our forest. $\mathrm{N_{leaf}}$ is a measure of the similarity between these two objects since objects that have a similar path inside the decision tree have similar features, and as a consequence are represented by the same model. We normalise $\mathrm{N_{leaf}}$ by dividing it by the number of trees in the forest which classify both objects as \emph{real}, $\mathrm{N_{trees}}$, giving us a similarity measure for every pair of objects. We define the distance between objects $i$ and $j$ as: 
$D_{ij} = 1 - \mathrm{N_{leaf}/N_{trees}}$. $D_{ij}$ can range from 0 (objects that ended up in the same leaf in all the trees), which is the distance of an object to itself, to 1 for objects that never ended up in the same leaf. 

We note that the standard way to measure distances between objects with RF is counting the number of tress in which the objects ended up in the same leaf, regardless of the leafs prediction (e.g., \citealt{shi06}), in this case $\mathrm{N_{trees}}$ is constant and equals to the number of trees in the forest. However, we find that the quality of the outliers improves when taking into account only leaves that predict both of the objects to be \emph{real}. This may be due to the fact that a model (the path in a decision tree) that describes a \emph{synthetic} object does not describe well a \emph{real} object and vice versa. The galaxies in our sample are classified as \emph{synthetic} by a decision tree with a probability of up to 25\%. This probability depends on the SNR of the galaxies, such that the probability to be classified as \emph{synthetic} increases with decreasing SNR. We note however that each decision tree has access to a limited number of features and therefore a galaxy with high SNR can be classified as \emph{synthetic} in some subset of the trees, thus adding noise to the distance measure. We further refer the reader to the work by \citet{shi06}, where they discuss different methods of synthetic data construction, distance assignment, and discuss the difference and similarity to the Euclidean distance.

Using this distance measure, we can cluster the objects but our purpose is to find outliers. The distance between all the pairs of objects is a matrix with dimensions 2\,355\,926 X 2\,355\,926, more than 20TB in double precision. Since we are interested in outlying objects, which have a large distance from most of the other objects in the dataset, it is enough to average the distances of a given object to all the others, resulting in a 2\,355\,926-sized vector that represents the `weirdness' score of each object. This vector is of a reasonable size and we can sort it and extract the objects with the highest weirdness score from the entire population -- the outliers. The sorted vector of averaged distances essentially orders the galaxies from the most mundane, with a weirdness score of 0, to the most unusual, with a score of 1. Therefore, the threshold for the outlying objects is user-defined. We also note that due to the randomness of the RF in its feature and object selection, the decision trees vary from one run to the next, and the weirdness score that we obtain is not unique. However, we find that when using a sufficient number of decision trees (see details in section \ref{s:para}) an object with a high weirdness score in one run will have a high weirdness score in different runs as well.

We measure the weirdness score for the illustration data shown in figure \ref{f:syn_data_illus} and show it in the right panel of figure \ref{f:weirdness_score_color}, where dark blue represents objects with high weirdness score and yellow represents objects with a low weirdness score. One can see that objects that are located in the tail of the distribution and noisy points tend to have higher weirdness scores. We show in figure \ref{f:dist_hist} a histogram of the weirdness score for all the galaxies in our sample. One can see that the distribution is smooth and decreases monotonically with the weirdness score. The presence of peaks in the histogram would have implied the existence of clusters of objects, composed of galaxies which have a short distance to the cluster members, but a long distance to galaxies in different clusters. The smoothness of the distribution that we measure may imply that the classifier is sensitive even to small differences between seemingly similar galaxies. This is essentially over-fitting, which is typically bad for supervised algorithms that are expected to generalise well to new measurements, but serves our purpose to find outliers. Alternatively, a smooth distribution in the averaged distance can happen if the galaxy spectra form a continuous sequence in spectral shape (e.g., \citealt{ascasibar11}). In Section \ref{s:outlier} we analyse the 400 weirdest objects, which lie to the right of the dashed blue line.  

\begin{figure}
\includegraphics[width=3.25in]{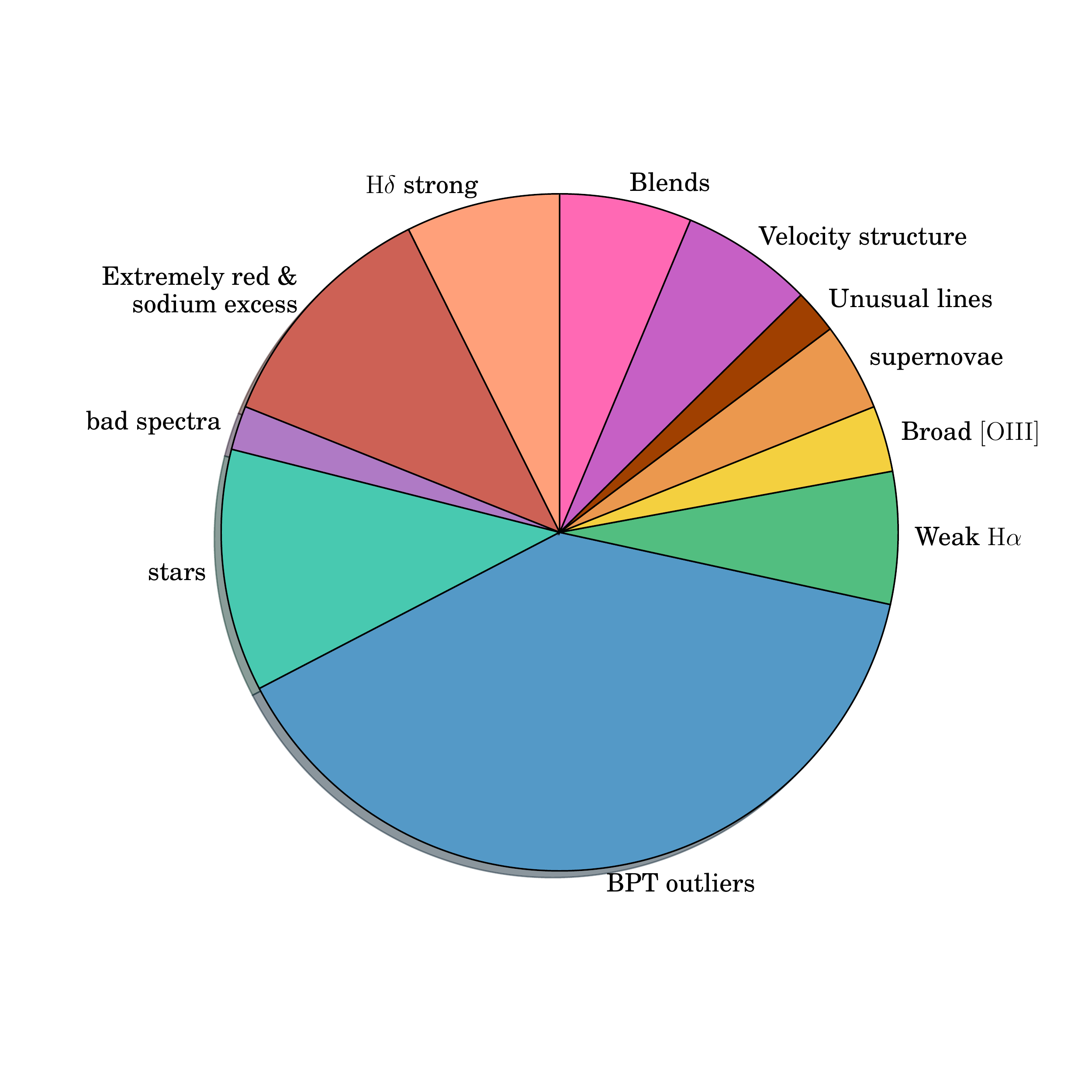}
\caption{Distributions of the classes of outlying objects we find. Clearly, we find a wide variety of phenomena. It is interesting that only a small fraction (2.5\%) are due to instrumental problems. 
}\label{f:pie_chart_objs}
\end{figure}

\begin{figure*}
\includegraphics[width=1\textwidth]{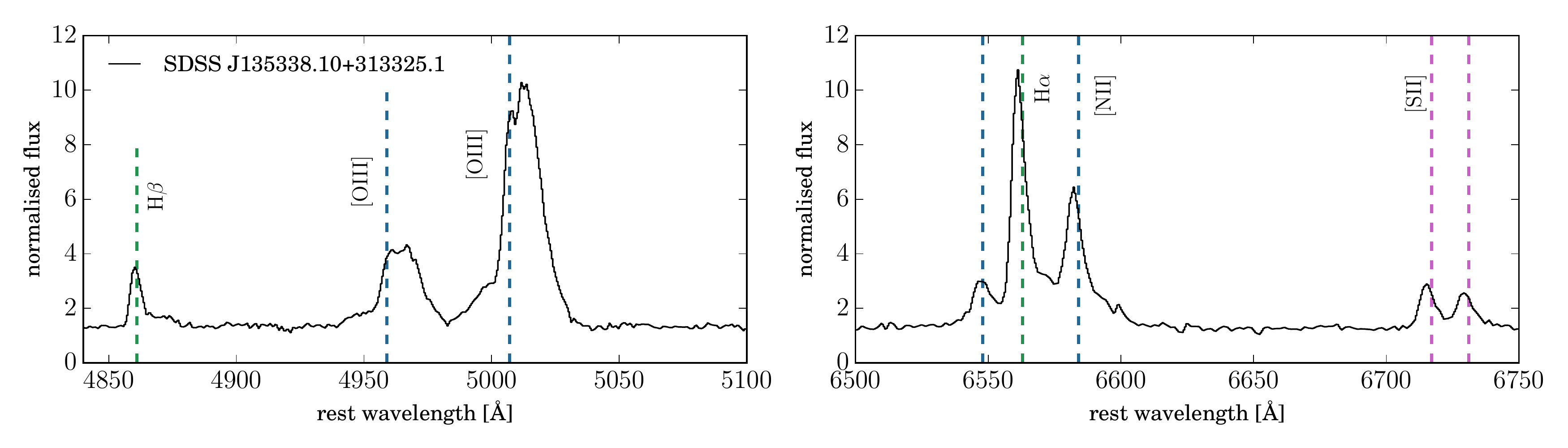}
\caption{SDSS J135338.10+313325.1 (2117-54115-230) has a number of velocity components in its Balmer and forbidden lines. There are two $\mathrm{[OIII]}$ emission peaks and another blue-shifted component, probably due to an outflow (left panel). In the Balmer emission lines one can see a broad component that is redshifted (see right panel). The vertical dashed lines mark the rest-frame central wavelength of emission lines, based on the redshift estimation of the SDSS pipeline.
}\label{f:nice_obj}
\end{figure*}

\subsection{Implementation}\label{s:para}
As noted above, RF is a naturally parallelisable algorithm due to the randomness of its object and feature selection. The size (2 $\times$ 2\,355\,926 objects) and complexity (15\,700 features) of our data prevent us from applying this algorithm with the entire data loaded to memory and the analysis to be done with a single process. Instead, we break the data into 8 bins of signal to noise ratio (SNR) of equal size, each bin contains roughly 300\,000 galaxy spectra. We define the SNR of a spectrum as the median SNR per pixel, as given by the SDSS pipeline. 

In addition to shorter running time, we find that dividing the spectra into SNR bins improves the quality of the outliers we find. We show the distribution of galaxy SNR in the SDSS in figure \ref{f:snr_distribution}, where one can see a smoothly declining distribution, except for a prominent peak at SNR$\sim$2.5 (with roughly 700\,000 galaxies). Like other machine-learning classifiers, RF builds a model that is mostly sensitive to the common object in the sample while objects that are rare are not well described. Therefore, when training without SNR bins, the decision trees are optimised to distinguish between \emph{real} and \emph{synthetic} data for a galaxy with SNR$\sim$2.5, but not for galaxies with higher SNR. In practice, the decision trees rank wavelengths with prominent emission lines such as $\mathrm{H\alpha}$ and $\mathrm{[OIII]}$ high in the feature importance ranking, while ignoring the weaker emission lines such as $\mathrm{H\beta}$, $\mathrm{[SII]}$, and $\mathrm{[OI]}$, and ignoring absorption lines, since the information of a galaxy with SNR$\sim$2.5 is mostly contained in these strong emission lines. We therefore divide the sample into SNR bins and indeed find that the feature importance of weak emission lines and absorption lines increases for bins of higher SNR, while prominent emission lines are ranked high in all of the bins. 

Within a given SNR bin, we build 50 redundant chunks which contain 10\,000 objects, and while the galaxy spectra within a given chunk are unique, a specific galaxy spectrum can be included in several different chunks. We construct 200 decision trees for each of these chunks, while using only a subset of the features in the construction of a single tree (a square root of the number of features is used for a given decision tree, which is a typical choice). Therefore, for a given SNR bin we build 10\,000 decision trees while using 125 features for each tree. We test different numbers of objects per chunk and find that for 5\,000 -- 20\,000 galaxies in a chunk the weirdness score distribution is the same, resulting in the same quality of outliers. However, when using 1\,000 -- 3\,000 galaxies per chunk the result is degraded (e.g., supernovae and stars do not always have a high weirdness score, and standard galaxies sometimes have a high weirdness score), and the distribution in weirdness score appears flatter. We also note that increasing the number of decision trees for a fixed number of observations generally results in a better and more general classifier but increases the training time substantially. For our dataset, we find that the distribution in weirdness score stabilises and barely changes once we reach 200 trees per 10\,000 objects. This process is identical to the RF training procedure we presented in section \ref{s:algo} and the construction of decision trees per chunk is executed in parallel on multiple computer nodes. We use the decision tree classifier and RF classifier implementation in scikit-learn\footnote{\tiny
scikit-learn.org/stable/modules/generated/sklearn.tree.DecisionTreeClassifier.html
scikit-learn.org/stable/modules/generated/sklearn.ensemble.RandomForestClassifier.html} \citep{scikitlearn}.

Once we have the decision trees from the parallel processes, we merge them into a forest. The phase of propagating an object through the trees and comparing its leafs to that of the other objects is done in parallel as well. The parallelisation in this case is object based -- we pair the reference object with all the other objects and compute the distance between the reference object to the entire sample and then we average it. These details were tailored to the data at hand, and to our available computational resources. We stress that they affect execution time but not our scientific results. In total the process takes 12 hours on 128 CPUs to obtain the weirdness score. The bulk of our computations was performed on the resources of the National Energy Research Scientific Computing Center\footnote{\texttt{http://www.nersc.gov/}}.

\subsection{Dimensionality reduction}\label{s:pca_fuya}
Dimensionality reduction and feature selection methods are often proposed when dealing with large-size and complex data. The goal of dimensionality reduction techniques is to extract the strongest, most prominent features in the data and use them for classification or regression problems. These methods assume that the data can be well represented in a lower dimensional space, thus allowing for faster and (sometimes) better statistical inference. A common method is Principal Component Analysis (PCA) which performs a linear decomposition of the data and provides eigenvectors and their strength coefficients for every object in the dataset (e.g. \citealt{zhang06, vandenberk06, rogers07, fiorentin07, bailey12}). When modelling a spectrum, the linearity of PCA may be a disadvantage due to the nonlinearity of spectral lines. Thus, it is not clear that applying PCA prior to the forest construction is beneficial for our algorithm. 

Nevertheless, we apply PCA decomposition on the 10\,000-sized chunks and use the coefficients provided by the decomposition as the features for the forest construction and outlier detection. We test different number of input coefficients: using 10\,000, 5\,000, 2\,000, 1\,000, 500, and 100. In all the cases the result suffers. Outlying objects tend to have a few coefficients with large (positive and negative) values, compared to the values for the same coefficients of standard objects. These large coefficients are not modelled well by the forest and tend to fall in the same leaf of a given decision tree, thus appearing similar to each other. Standard spectra which are intrinsically alike also tend to fall in the same leaf and appear similar. These two groups (intrinsically similar objects and outlying objects) show the same summed distances and therefore cannot be separated with a distance criterion. So while PCA might be advantageous for a classification problem with RF, by mitigating over-fitting, it impairs our weirdness measure. 

There are dimensionality reduction algorithms which are not linear. \citet{richards09} apply diffusion maps to galaxy spectra for data parametrisation and dimensionality reduction prior to using the spectra in a regression problem. \citet{vanderplas09} and \citet{daniel11} use Locally Linear Embedding (LLE) as a nonlinear dimensionality reduction technique for classification of SDSS spectra. Although these methods show great promise in the dimensionality reduction field, we chose not to use them since RF, contrary to many ML algorithms, deals with large number of features well, and applying these methods is expensive both in terms of running time and available memory. Furthermore, it is not trivial to apply these methods on millions of objects in the DFS framework and on data streams. More importantly, as noted above, the better they work, the more they are likely to harm our algorithm for outlier detection. 

\subsection{Comparison with other outlier detection algorithms}\label{s:other_algs}
There are several outlier detection algorithms that are based on unsupervised learning which are widely used in various fields. These algorithms can be divided into two types. The first type is distance-based anomaly detection, in which a distance is computed between every pair of objects in the sample and outliers are defined as objects that are distant from most of the objects in the sample. This includes Euclidian-based distance assignment, PCA-decomposition-based distance assignment, and unsupervised RF distance assignment. The running time of these algorithms scales as $\mathrm{O(N^{2})}$, where $\mathrm{N}$ is the number of objects in the sample, since a pairwise distance must be computed. The second type of algorithms focus on isolating outliers rather than modelling the entire data and defining outliers as objects that deviate from this model. The most popular such algorithm is Isolation Forest (iForest; \citealt{liu08}). Such algorithms are typically linear in time. We refer the reader to \citet{liu08} and \citet{goix16} for comparison between different algorithms for different data-sets.

We compare the quality of the outliers we find with the following algorithms: One-Class SVM \citep{scholkopf99}, iForest \citep{liu08}, standard unsupervised RF \citep{shi06}, and our method introduced here. While our method is distance based, it is not as fast as iForest for example, but it is naturally parallelisable, reducing wall-time. We gauge the quality of the outliers we find with the various methods qualitatively, using domain knowledge. For example, we expect supernovae and stars to have high weirdness scores, and galaxies that are only brighter but otherwise normal to have low scores. We also compare the number of false-positives found by these algorithms, that is the number of standard-looking objects that are marked as outliers by the algorithms. 

Visually inspecting the 400 weirdest galaxies found by each of these algorithms, we find that our method significantly outperforms the others, finding essentially only interesting galaxies, as thoroughly discussed below. The standard unsupervised RF algorithm tends to pick a larger number of false-positives. For One-Class SVM and iForest the galaxies that are found as outliers include mostly galaxies with residual cosmic rays, galaxies with extremely strong emission lines, stars, and a few false positives. Generally, we find that these algorithms detect objects that show outlying properties in specific features (i.e., wavelengths) but are not sensitive to outlying relation between features such as extremely reddened galaxies, outlying line ratios, and unusual emission lines. Although the former are indeed outliers in a statistical sense, the covariance-based outliers are more interesting and offer the opportunity to discover subtle changes that represent an outlying physical behaviour. 

\begin{figure*}
\includegraphics[width=1\textwidth]{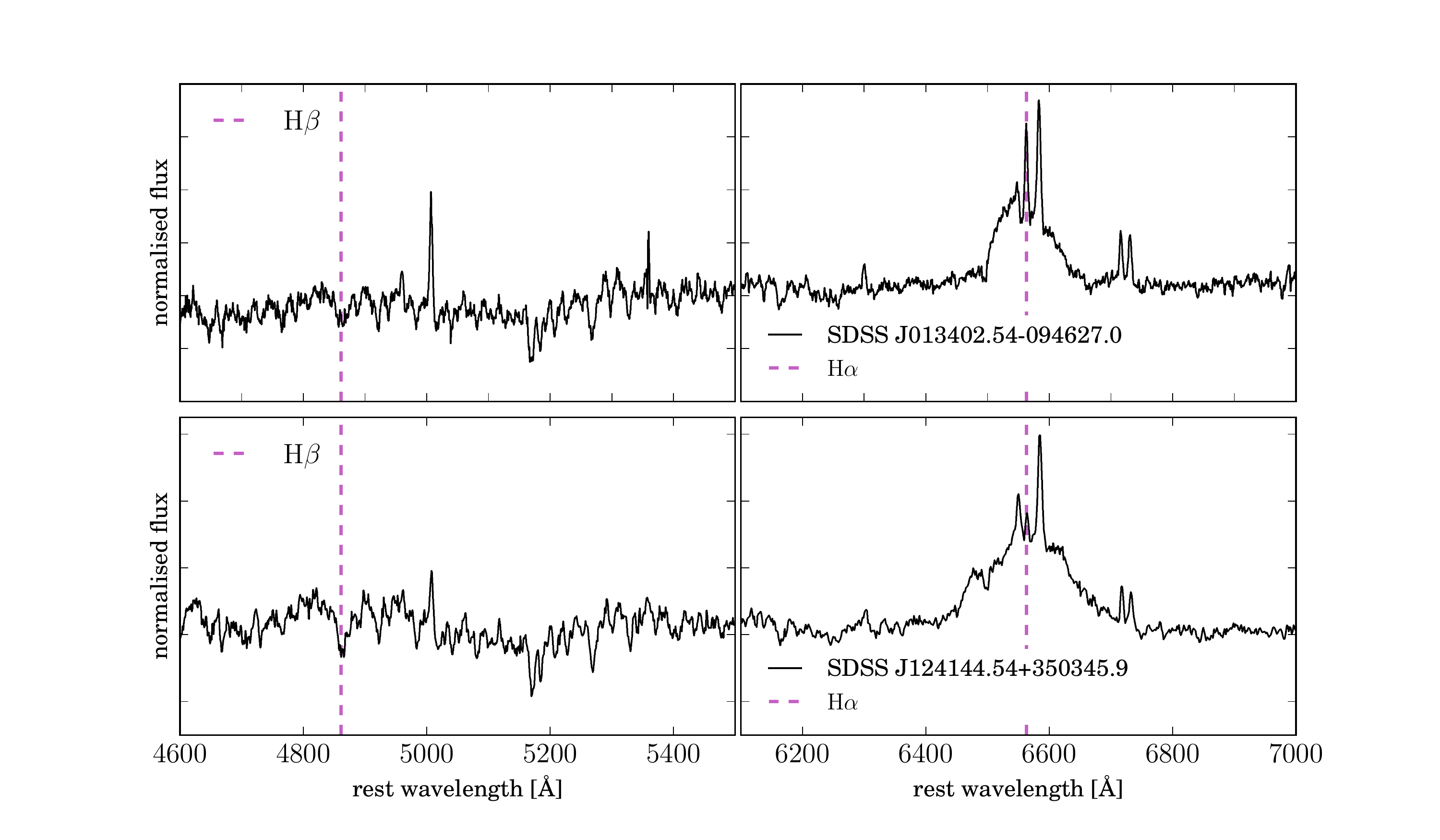}
\caption{
Two examples of galaxies with an additional structure near the H↵-[NII] complex. SDSS J013402.54-094627.0 (663-52145-306; top panels) has an unusual asymmetric emission near $\mathrm{H\alpha}$ and no emission near $\mathrm{H\beta}$. SDSS J124144.54+350345.9 (2022-53827-286; bottom panels) has a more symmetric structure, with a centroid that is slightly shifted from the $\mathrm{H\alpha}$ wavelength. In both of the cases we find similar structure in $\mathrm{H\beta}$ after we subtract the best stellar population synthesis model, though we do not detect continuum emission from an accretion disk.}
\label{f:halpha_bumps}
\end{figure*}

\section{Outlying Galaxies}\label{s:outlier}
We systematically inspect the 400 weirdest galaxies that we find. We caution the reader that their relative numbers are not very meaningful, since our choice to study 400 is arbitrary, and limited by the time necessary to inspect all the aspects of each galaxy, as we do below. This includes visual inspection of the SDSS spectrum and imaging, line measurements where applicable, a literature search (since this sample was so extensively studied), and other more specific measurements for specific cases. There are undoubtedly many more interesting sources with a marginally lower weirdness score. We encourage our readers to explore further. A table with a weirdness score for all the 2M galaxies is available at \texttt{http://www.wise-obs.tau.ac.il/$\sim$dovip/weird-galaxies/}, and an IPython notebook with the algorithm is available at \texttt{https://github.com/dalya/WeirdestGalaxies}.

Out of 400, we call 10 objects `bad spectra' and they consist of objects with calibration problems and sky contamination, we find 47 stars classified as galaxies by the SDSS pipeline. 27 objects we call `blends', which includes galaxy-galaxy lenses, chance alignment of two objects, and multiple emission line systems. We find 186 galaxies with unusual emission-line ratios. 13 galaxies sport an unusually broad [OIII] emission, in fact, it is so broad in some cases that the SDSS pipeline mistakes it for a broad $\mathrm{H\alpha}$ emission line at a wrong redshift. We also find 33 galaxies with extreme Balmer absorption up to $\mathrm{H\delta}$ and beyond, and 38 objects with an excess of NaID absorption. 18 outliers are galaxies that host supernovae, 9 show unusual emission lines (e.g., coronal lines, $\mathrm{[NI]}$). We find 41 galaxies with unusual velocity structures -- 11 of them show an additional structure around the $\mathrm{H\alpha}$-$\mathrm{[NII]}$ complex, and 13 show double-peaked emission lines. We show in figure \ref{f:pie_chart_objs} a pie chart that summarises all the outlying objects and their classification. Out of the 400 outliers, only 74 outliers were previously reported by other studies -- in all of these cases the authors have constructed a specific algorithm that is tailored exactly to find the kind of objects they are looking for. We list all the outlying galaxies in appendix \ref{a:app} with classification and references to previous studies when relevant. Throughout this section, we note specific galaxies with their SDSS name (e.g., SDSS J131027.46+182617.4) and their SDSS PLATE-MJD-FIBER. 

\subsection{Unusual velocity structure}\label{s:vel}
We find 42 galaxies with an unusual velocity structure in their emission lines. Out of these, 11 galaxies show an additional structure around the $\mathrm{H\alpha}$-$\mathrm{[NII]}$ complex, 13 galaxies have double-peaked emission lines, three show two broad components in the $\mathrm{H\alpha}$ and $\mathrm{H\beta}$. We also find 13 galaxies with unusually broad $\mathrm{[OIII]}$ emission, and the remaining five galaxies show a number of interesting velocity components in their emission lines. We list these five objects in table \ref{table:unus_vel}. 

We present in figure \ref{f:nice_obj} one of the four galaxies that shows the most complex velocity structure as an example. SDSS J135338.10+313325.1 (2117-54115-230) has a number of velocity components in its Balmer and forbidden lines. We mark the rest-frame wavelengths of the emission lines based on the redshift estimation of the SDSS. One can see that this galaxy has two $\mathrm{[OIII]}$ emission peaks and another blue-shifted component, probably due to an outflow. In the Balmer emission lines one can see a broad component that is redshifted compared to the central rest-frame wavelength (more prominent in the $\mathrm{H\alpha}$ emission). The galaxy seems disturbed in the SDSS imaging, and this velocity structure could arise from asymmetries near the centre of the galaxy which is picked up by the spectroscopic fibre. We also note SDSS J113818.38+060620.1 (4767-55946-889), which has an ultra-luminous, galaxy-wide narrow line region and is studied by \citet{schirmer13} as it is a "green bean". \citet{schirmer13} construct a sample of 29 such galaxies from SDSS DR8, and claim them to be among the rarest objects in the universe. 

\subsubsection{Additional structure near $\mathrm{H\alpha}$}\label{ss:halpha_bumps}
We find 11 objects with an additional structure near the $\mathrm{H\alpha}$-$\mathrm{[NII]}$ complex. The stellar continuum contribution in these galaxies is stronger than nebular emission, mostly near $\mathrm{H\beta}$, thus similar features are not easily detected in $\mathrm{H\beta}$ emission. In order to isolate the emission of the gas from the stellar component, we subtract the best fitting population synthesis model for each of the galaxies. We use pPXF \citep{cappellari12}, which is a public code that extracts the stellar kinematics and stellar population from absorption-line spectra of galaxies \citep{cappellari04}. The code uses the MILES library, which contains single stellar population synthesis models from \citet{vazdekis10}. In several galaxies, we obtain unrealistic stellar population fits due to the broad $\mathrm{H\alpha}$ emission. We therefore ignore pixels that are dominated by the broad emission features prior to the fitting, and obtain good fits. 

\begin{figure} 
\includegraphics[width=3.25in]{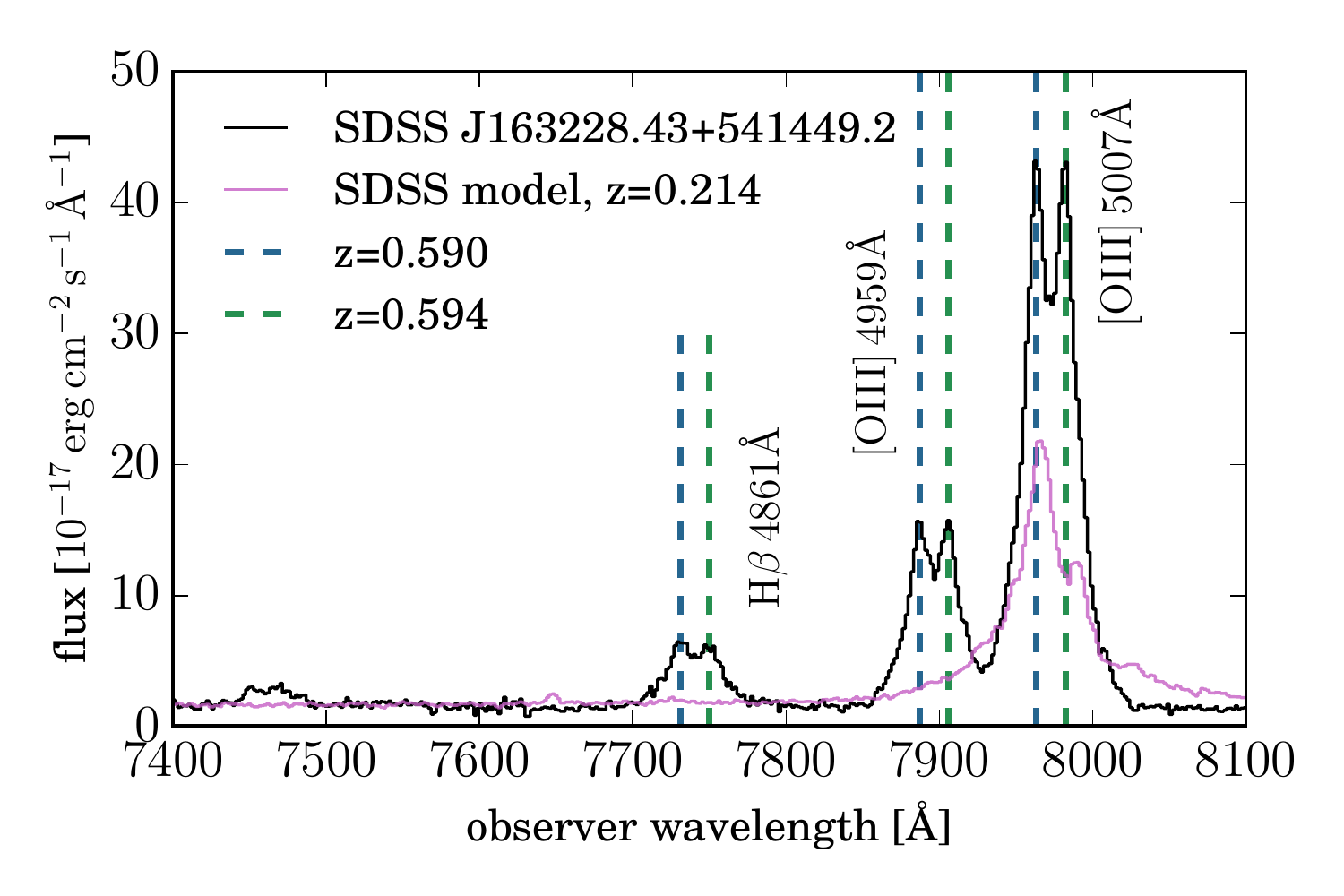}
\caption{
Example of an object with double-peaked emission lines in the observer frame (SDSS J163228.43+541449.2; 6788-56428-257). The object shows two strong sets of emission lines which are separated by 700 $\mathrm{km\,s^{-1}}$ (black line). The SDSS pipeline estimate the redshift of the object to be $z=0.214$, fitting a model where a broad $\mathrm{H\alpha}$ component accounts for the wide and double $\mathrm{[OIII]}$ line (SDSS model in pink). We estimate the redshift of the emission line components to be $z=0.590$ and $z=0.594$, and mark the $\mathrm{[OIII]}$ and $\mathrm{H\beta}$ emission line wavelengths in the observers frame according to our redshift estimation. The galaxy shows extremely symmetric components, compared to other double-peaked emission line galaxies, which may indicate that we are seeing a disk-like structure. 
}\label{f:double_peaked_example}
\end{figure}

The emission spectra of these galaxies show a wide range of properties. For some of the galaxies, we detect a similar emission structure around $\mathrm{H\beta}$, suggesting that the broad features are due to Balmer emission, and in some of these the broad emission is not centred around the central wavelength of the narrow emission, suggesting an outflow or a complex geometry (e.g., SDSS J154327.02+075442.1, 1724-53859-116, which shows an outflow in the Balmer emission lines). We find that several galaxies show diagonally-shaped continua (see section \ref{ss:red}), which suggests that the broad $\mathrm{H\beta}$ emission is absorbed by dust reddening and is not detected due to our limited SNR. Furthermore, we find that while for five of the galaxies the additional structure can be attributed to a broad $\mathrm{H\alpha}$ emission, or to two $\mathrm{H\alpha}$ velocity components which are blended, the remaining six show asymmetric structures which cannot be described by a broad Gaussian. We list these objects in table \ref{table:halpha_bumps}. 

\begin{figure*}
\includegraphics[width=0.34\textwidth]{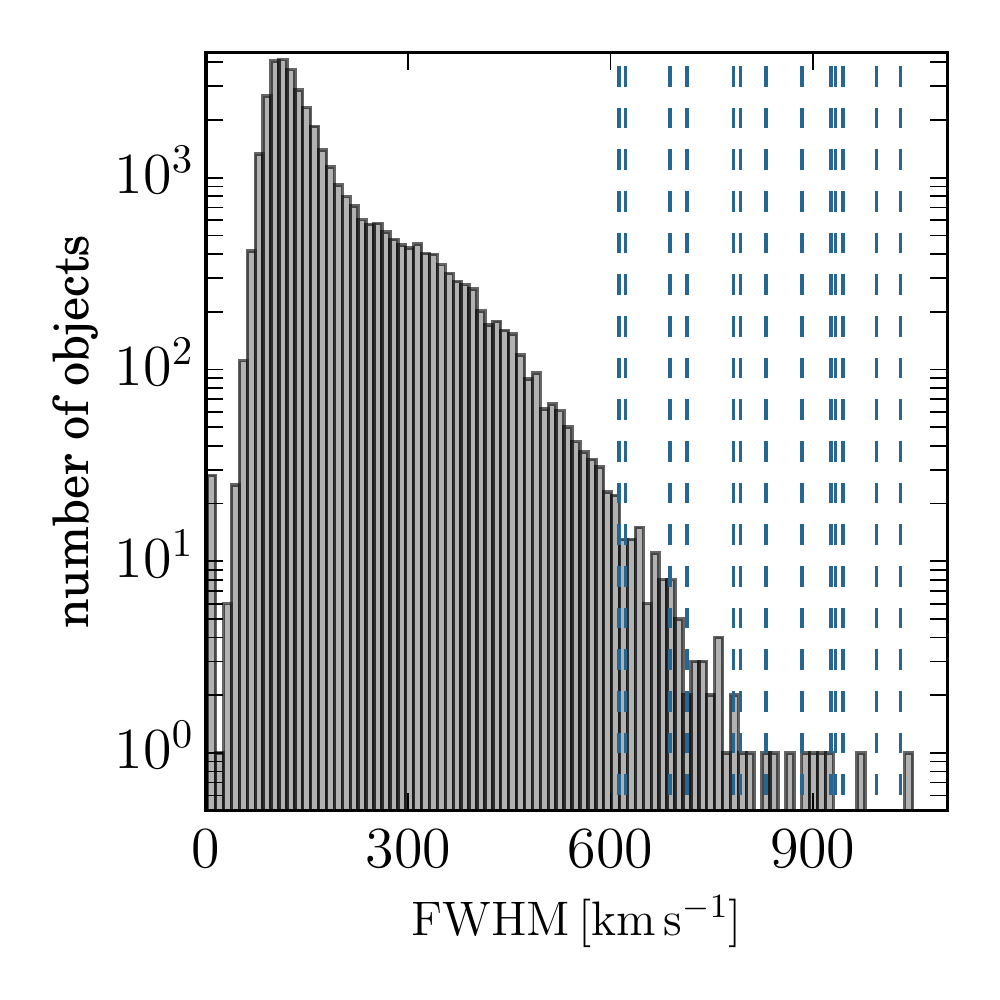}
\includegraphics[width=0.60\textwidth]{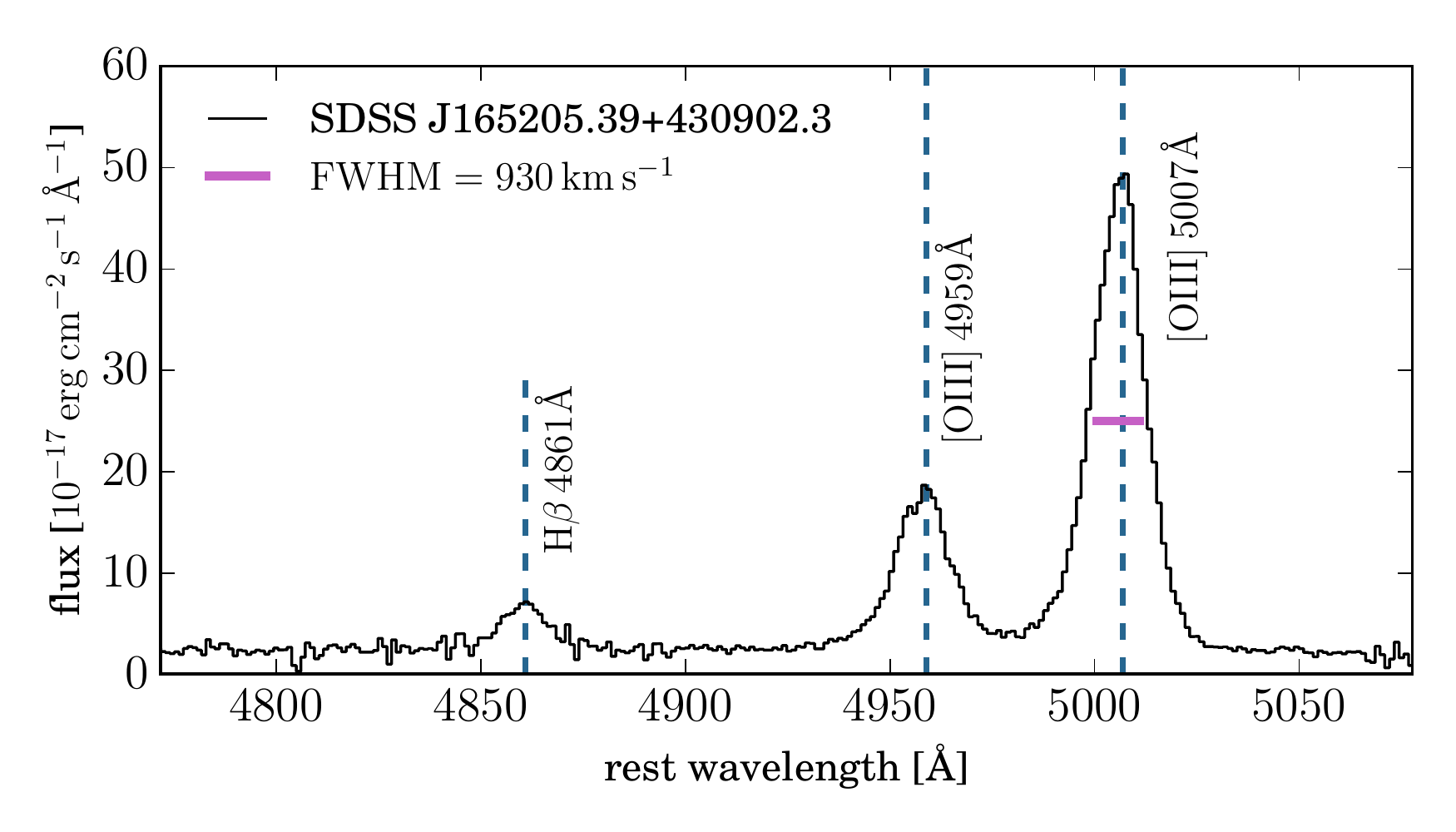}
\caption{ Histogram of the FWHM of the forbidden emission lines of galaxies from the MPA-JHU catalogue (left panel), compared to the FWHM of the $\mathrm{[OIII]}$5007\AA\, emission line we measure for our objects. One can see that our objects reside in the tail of the width distribution. As an example, we show SDSS J165205.39+430902.3 (6031-56091-0007; right panel; black), $z=0.634$, where $\mathrm{H\beta}$, $\mathrm{[OIII]}$4959\AA, and $\mathrm{[OIII]}$5007\AA are marked with blue dashed lines. The FWHM of $\mathrm{[OIII]}$5007\AA\, is $930\,\mathrm{km\,s^{-1}}$ (pink).
}\label{f:broad_oiii_example}
\end{figure*}

We present two examples of such objects in figure \ref{f:halpha_bumps}. SDSS J013402.54-094627.0 (0663-52145-0306; top panels) shows an asymmetric profile, which cannot be well represented by a single or double Gaussian and is not centred around the $\mathrm{H\alpha}$ emission line. After subtracting the best stellar population synthesis model from the spectrum, we detect two broad emission components around $\mathrm{H\beta}$, which suggests that the broad feature around  $\mathrm{H\alpha}$ is due to Balmer emission (rather than $\mathrm{[NII]}$ emission or a mix of the two). However, using two broad emission components for $\mathrm{H\alpha}$ (and all the narrow emission lines) we are not able to reproduce the shape of the emission we observe. We find that adding an additional broad component in absorption can reproduce the profile we observe. Broad Balmer absorption is typically seen in Type II supernovae ejecta (see for example \citealt{faran14}) or extremely dense outflows in high redshift quasars (see \citealt{williams16} and references therein). An alternative interpretation for this profile is a circular, relativistic, Keplerian disk which is observed in a small fraction of AGN \citep{eracleous03}. SDSS J124144.54+350345.9 (2022-53827-0286; bottom panels of figure \ref{f:halpha_bumps}) has a more symmetric structure, only slightly shifted from the $\mathrm{H\alpha}$ wavelength. We detect the same structure in $\mathrm{H\beta}$ after subtracting the stellar population contribution. In both of these cases, we do not detect continuum emission from an accretion disk.

\subsubsection{Double-peaked emission line galaxies}\label{s:doublep}
Another subset of objects consists of galaxies with double peaked emission lines, of which we find 13, 10 of which show narrow and 3 show broad double-peaked emission lines. Previous studies conducted searches in the SDSS spectroscopic data with the purpose of finding double peaked emission lines in galaxies and AGN up to the seventh data release (DR7; \citealt{wang09}; \citealt{liu10}; \citealt{smith10}; \citealt{ge12}; \citealt{pilyugin12} and references therein). \citet{smith10} searched for binary AGN and by visual inspection of SDSS AGN spectra found 148 spectra with a double peaked $\mathrm{[OIII]}$ profile. \citet{pilyugin12} looked for galaxies with two simultaneous starbursts and by visual inspection of the SDSS spectra found a few hundreds of galaxies with double peaked emission lines, and studied 55 of them, which had a reliable two component decomposition. \citet{ge12} compiled the first systematic sample of 3030 double peaked emission lines objects.

Out of the 10 galaxies with double-peaked narrow emission lines, two were reported by previous studies. SDSS J103007.07+412353.5 (1360-53033-0186) is part of the sample presented by \citet{pilyugin12} and \citet{ge12}. SDSS J131642.90+175332.6 (5867-56034-0304) is discussed specifically by \citet{xu09}, where the authors suggested and discussed a few scenarios which can potentially account for the double peaked profiles of all of the lines, such as a binary AGN, galaxy merger, and jet-cloud interaction. This object is of a particular interest since it shows double emission profiles in all its nebular lines and an additional broad component (FWHM of $1400\,\mathrm{km\,s^{-1}}$) in most of its forbidden lines. Two additional objects (out of the 10) are worth noting since the SDSS pipeline failed to estimate their redshift correctly: SDSS J223623.78+112824.0 (5049-56103-0273) and J163228.43+541449.2 (6788-56428-0257). The velocity separation between the two peaks is $550\,\mathrm{km\,s^{-1}}$ and $710\,\mathrm{km\,s^{-1}}$, higher compared to the velocity separation of double-peaked galaxies (see velocity separation distribution in \citealt{pilyugin12}). We present, as an example, the object SDSS J163228.43+541449.2 in figure \ref{f:double_peaked_example} (black) compared to the best fitting model of the SDSS pipeline (pink) which mistakes the redshift to be $z=0.214$, identifying the broad $\mathrm{[OIII]}$ line as $\mathrm{H\alpha}$. We estimate the redshift of the two emission line components to be $z=0.590$ and $z=0.594$ with a velocity separation of $710\,\mathrm{km\,s^{-1}}$. These galaxies were not reported in previous studies.

We find 3 galaxies which show double-peaked broad emission lines, all of them are reported by previous studies as well. Due to this structure, the SDSS pipeline fails to fit a correct redshift to two of them, fitting the broad double-peaked $\mathrm{H\alpha}$ emission as $\mathrm{Ly\alpha}$ emission. For SDSS J013253.31-095239.3 (662-52147-180), the SDSS estimate the redshift to be 5.73, while we measure it to be 0.259. The broad emission components belong probably to $\mathrm{H\alpha}$ emission (and not the $\mathrm{[NII]}$) since we observe the same structure in $\mathrm{H\beta}$. These components are located at a velocity of 2880 $\mathrm{km\,s^{-1}}$ (blue-shifted) and -4890 $\mathrm{km\,s^{-1}}$ (redshifted). SDSS J114051.58+054631.1 (838-52378-460) also show two broad emission lines, offset by 4600 $\mathrm{km\,s^{-1}}$ and -5800 $\mathrm{km\,s^{-1}}$, and we estimate its redshift to be 0.131. For additional information, we refer the reader to the studies that discovered and studied these objects: \citet{strateva03}, \citet{wu04}, \citet{bian07}, and \citet{strateva08}. We list all the double-peaked galaxies in table \ref{table:double_peaked}.

\subsubsection{Broad $\mathrm{[OIII]}$ emission line}\label{ss:oiii_broad}
We find 13 galaxies with relatively broad $\mathrm{[OIII]}$ emission lines. For 4 of these objects the SDSS pipeline fails to fit a model with a correct redshift, identifying the unusual $\mathrm{[OIII]}$ as $\mathrm{H\alpha}$. We derive the redshift of the objects by fitting the $\mathrm{H\beta}$, $\mathrm{[OIII]}$4959\AA, $\mathrm{[OIII]}$5007\AA\, complex, with just three Gaussians, and extract the FWHM of the $\mathrm{[OIII]}$ 5007\AA\, line. The values we derive range from $600\,\mathrm{km\,s^{-1}}$ to $1000\,\mathrm{km\,s^{-1}}$.

In order to compare the FWHM of our $\mathrm{[OIII]}$5007\AA\, emission line to that of the general population, we use the MPA-JHU catalogue, which contains emission line measurements of 927\,552 galaxies from SDSS DR7 \citep{kauffmann03a, brinchmann04, tremonti04}. They also derived the velocity of all the forbidden lines (they constrained the velocity to be the same in all the forbidden lines in their fitting process) and converted it to FWHM value, assuming a Gaussian profile. We use galaxies for which the velocity measurement is at least three times higher than its uncertainty, thus rejecting low SNR objects. We show in the left panel of figure \ref{f:broad_oiii_example} the histogram of the FWHM of the entire galaxy population compared to the FWHM of our objects. One can see that these objects are located on the tail of the distribution of the entire population. As an example, we present SDSS J165205.39+430902.3 (6031-56091-0007) in the right panel (black). The redshift derived by the SDSS is $z=0.244$ while the redshift we derive is $z=0.634$, and we measure a FWHM of $930\,\mathrm{km\,s^{-1}}$ for the $\mathrm{[OIII]}$ emission line. We note that in the case of broad $\mathrm{[OIII]}$ emission, samples constructed from the SDSS can be incomplete and biased due to incorrect redshift estimation. At a redshift of $z\sim0.5$, where $\mathrm{H\alpha}$ is outside the SDSS wavelength range, broad $\mathrm{[OIII]}$ emission can be confused with broad $\mathrm{H\alpha}$ and $\mathrm{Ly\alpha}$ emission, and indeed there are hundreds of quasars with incorrect redshift estimation in the SDSS database (see \citealt{reyes08} and \citealt{yuan16}). We list the broad $\mathrm{[OIII]}$ emission line galaxies in table \ref{table:broad_oiii}.

\begin{figure*}
\includegraphics[width=0.6\textwidth]{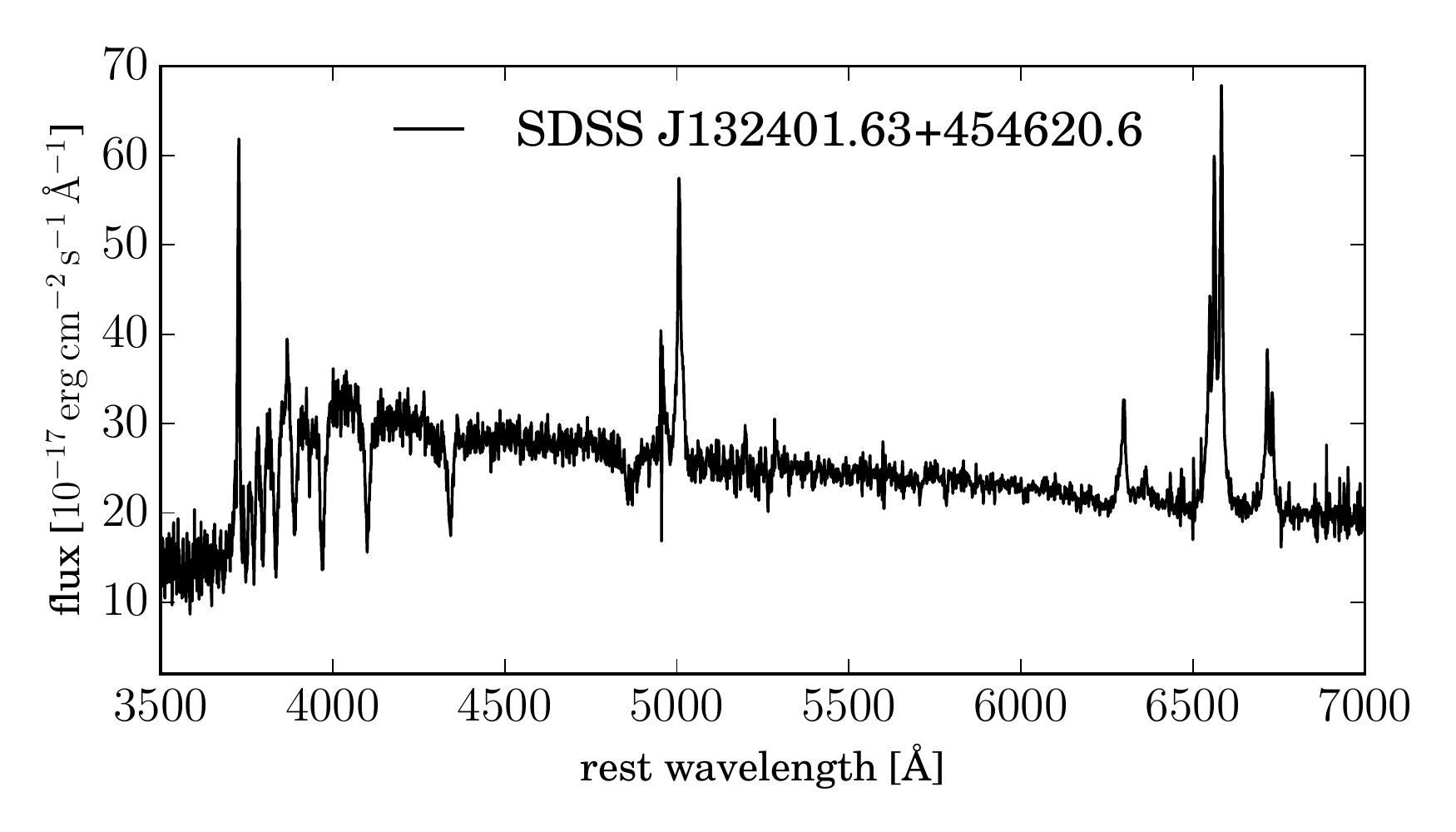}
\includegraphics[width=0.34\textwidth]{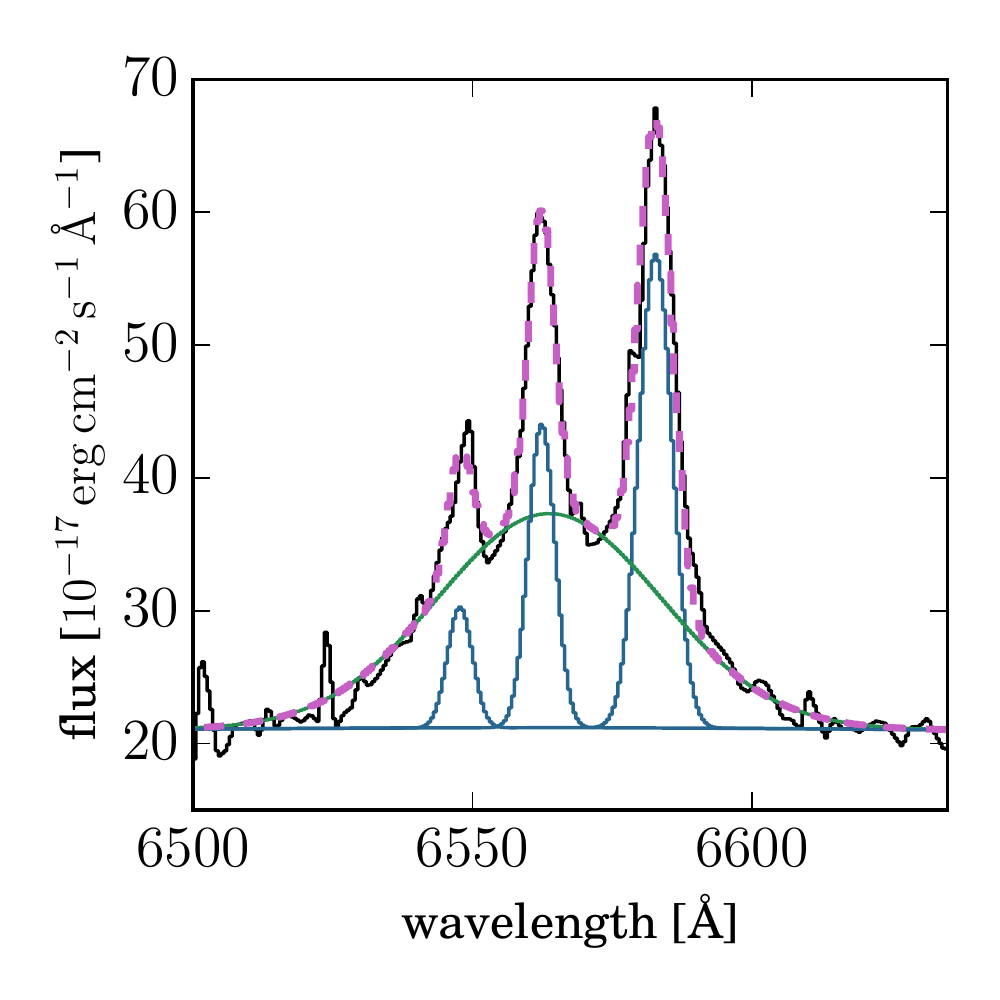}
\caption{SDSS J132401.63+454620.6 (1463-53063-262) -- $\mathrm{H\delta}$ strong galaxy which also has a broad $\mathrm{H\alpha}$ emission line (left panel). This object is peculiar since the strength of the $\mathrm{H\delta}$ indicates a recent change in its star formation activity, however an AGN is still detectable through broad $\mathrm{H\alpha}$ emission. We fit the lines (right panel), find that the FWHM of the broad $\mathrm{H\alpha}$ component is $2150\,\mathrm{km\,s^{-1}}$, and that the line ratio $\mathrm{log\,[NII]/H\alpha}=0.28$ is consistent with gas that is exposed to hard X-ray radiation.
}\label{f:hdelta_AGN}
\end{figure*}

In many galaxies, the $\mathrm{[OIII]}$ emission lines show an additional blue-shifted component, which is usually interpreted as being due to outflows. Forbidden lines are used to study the kinematics of galaxies, since in most of the cases they have narrow resolved profiles (unlike the $\mathrm{H\alpha}$ and $\mathrm{H\beta}$ for example). As a consequence, different velocity components can be separated and studied. However, broad $\mathrm{[OIII]}$ lines that can be well described by a single Gaussian are rare (e.g., \citealt{alexander10, greene11, harrison14, carniani15, zakamska16}). 

\subsection{E+A galaxies}\label{ss:hdelta}
E+A galaxies (also known as k+a galaxies, `post-starburst', and $\mathrm{H\delta}$-strong galaxies) are galaxies with strong Balmer $\mathrm{H\delta}$ absorption in their spectra, with values typically larger than 4--5\AA. The emission in these galaxies is dominated by A-stars, with lifetimes of around one Gigayear. Therefore, stars were formed recently, O-stars and B-stars have all evolved past the main-sequence, but A-stars have not yet. Consequently, E+A galaxies must have undergone a recent (and quite abrupt) change in their star formation. These galaxies comprise 3\% of the general galaxy population \citep{goto03} and in many of them it appears that a significant fraction of the galaxy stellar mass was created in the starburst (20-–60\%). A special sub-class of these are galaxies that do not show $\mathrm{H\alpha}$ and $\mathrm{[OII]}$ emission lines, indicating that these galaxies are not currently star forming, but have undergone a recent starburst. These galaxies are  observed during a short-lived phase in their evolution, and are fewer than 1\% of the galaxy population \citep{goto03, goto04, balogh05, goto07, melnick13, alatalo16}.

Out of 400 outlying galaxies, we identify strong $\mathrm{H\delta}$ absorption in 33, 28 of which were discussed in previous studies of such galaxies (see references above). We look for $\mathrm{H\alpha}$ and $\mathrm{[OII]}$ emission in our objects and find that only 7 out of 33 show emission lines that indicate ongoing star formation, while 26 do not. We list the $\mathrm{H\delta}$ strong galaxies we find in table \ref{table:ea_galaxies}.

We note specifically SDSS J132401.63+454620.6 (1463-53063-262), which shows strong $\mathrm{H\delta}$ absorption and a broad emission components near the $\mathrm{H\alpha}$ emission line. We show the object in the left panel of figure \ref{f:hdelta_AGN}. We start by fitting a stellar population synthesis model to account for the stellar continuum, like in Section 4.1.1. We find a good fit with a population of stars of age 0.05--0.1 Gigayear, as expected for E+A galaxies. We then subtract the stellar contribution in order to study the properties of the emitting gas.

We start by fitting a single component to the narrow $\mathrm{H\alpha}$ emission and two additional narrow components to the $\mathrm{[NII]}$ emission and find that these cannot account for the entire emission. We therefore fit an additional, broad, component that is centred around the $\mathrm{H\alpha}$ central wavelength. The best fitting profile is presented in the right panel of figure \ref{f:hdelta_AGN}. The FWHM we obtain for the broad $\mathrm{H\alpha}$ is $2150\,\mathrm{km\,s^{-1}}$. We also measure the line ratio $\mathrm{log\,[NII]/H\alpha}$ for the narrow components and find it to be 0.28, consistent with gas that is exposed to hard X-ray radiation. We note however that the forbidden lines $\mathrm{[SII]}$, $\mathrm{[OI]}$, and $\mathrm{[OIII]}$ show weak broad features that are marginally detected, thus it is possible that the broad emission around $\mathrm{H\alpha}$ is composed of both $\mathrm{H\alpha}$ and $\mathrm{[NII]}$ contributions.

It is clear that this object is of special interest. The stellar composition of the galaxy suggests that star formation has started simultaneously in the entire galaxy (due to the narrow age distribution that we find) and that the galaxy has undergone a recent shut-down of its star formation (i.e., no O-stars and B-stars). On the other hand, we measure broad emission components that may suggest either AGN heating or shocked gas due to recent merger or nuclear outflows. In terms of the cosmic cycle (e.g., \citealt{hopkins06}), this galaxy may represent a very short-lived phase in general galaxy evolution, where the star formation had already terminated due to AGN feedback though enough gas is still left to fuel the accretion disk (see for example \citealt{springel05} and \citealt{netzer09}). Studying this galaxy and similar ones may allow one to study the feedback mechanisms that terminate star formation, which we do in Baron et. al. (in prep).

\begin{figure*}
\includegraphics[width=1\textwidth]{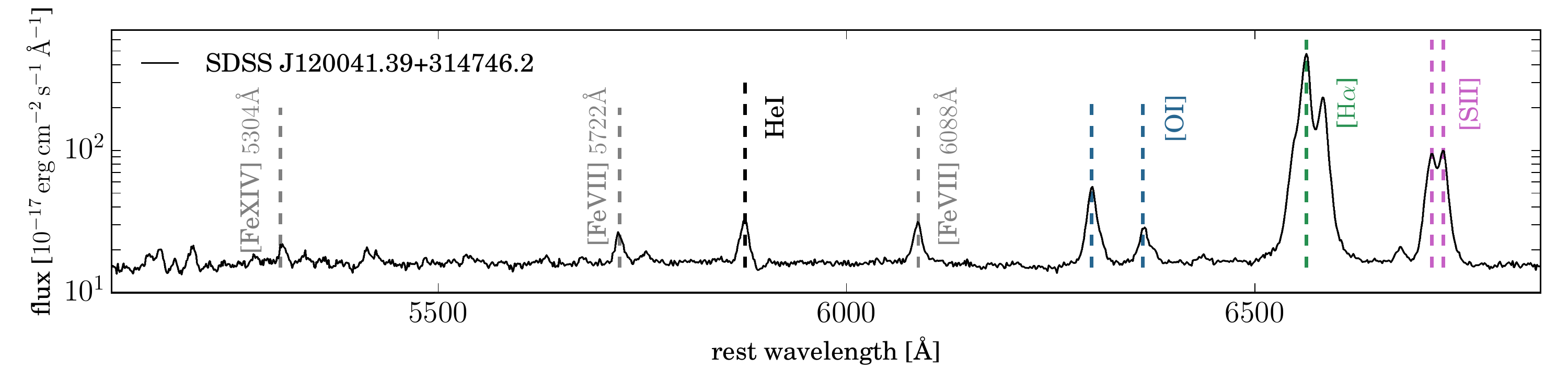}
\caption{Galaxy with detected high ionisation emission lines. SDSS J120041.39+314746.2 (6473-56363-672) is a galaxy at redshift $z=0.116$ for which we detect the coronal lines: [FeX] 6376\AA, [FeVII] 6088\AA, [FeVII] 5722\AA, [FeXIV] 5304\AA, and [FeVII] 3759\AA. This galaxy is not reported by \citet{wang12} since it is classified as AGN from its line diagnostics.}\label{f:CLE_example_plot}
\end{figure*}

\begin{figure*}
\includegraphics[width=1\textwidth]{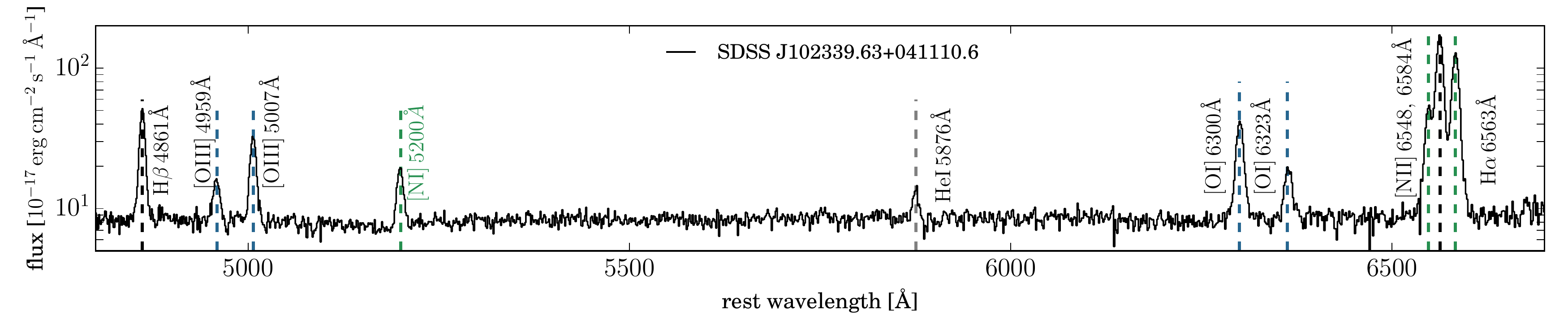}
\caption{An example of a galaxy with unusually strong $\mathrm{[NI]}$ emission. SDSS J102339.63+041110.6 (575-52319-200) is a galaxy at redshift $z=0.290$ and it is a member of a galaxy cluster, and it is detected in radio emission. We measure the ratio $\mathrm{[NI]/H\beta}=0.33$ in the spectrum, while in planetary nebulae the $\mathrm{[NI]}$ emission is 10--100 times weaker than $\mathrm{H\beta}$.}\label{f:ni_single_plot}
\end{figure*}

\subsection{Unusual emission lines}\label{ss:tdes}
We find 3 galaxies that show coronal lines in their spectra, and 10 galaxies with strong $\mathrm{[NI]}$\,5200\AA\, emission. We list these galaxies in table \ref{table:cls}.

\subsubsection{Coronal line emitters}
High-ionisation, or coronal lines arise in gas that is exposed to soft X-ray radiation. Several studies have suggested a connection between extreme coronal line emitters (ECLEs) and tidal disruption events (TDE), in which a star is tidally disrupted as it enters the tidal radius of a massive black hole \citep{komossa08, wang11, wang12}. \citet{wang12} were the first to conduct a systematic search for TDEs in the SDSS spectroscopic data by looking for ECLEs. They detected ECLEs by either looking for extreme coronal lines (at least one line is at least 20\% as strong as the $\mathrm{[OIII]}$ line) or coronal lines from a galaxy that is classified as an HII region on the BPT diagram. The authors found 7 objects in total that satisfy the criteria, one of them is found as an outlier by our algorithm (SDSS J093801.63+135317.0; 2580-54092-0470). It is worth noting that out of the 7 objects in their sample, only 4 objects are classified as galaxies by the SDSS pipeline (rather than QSO) and enter our sample.

\begin{figure*}
\includegraphics[width=1\textwidth]{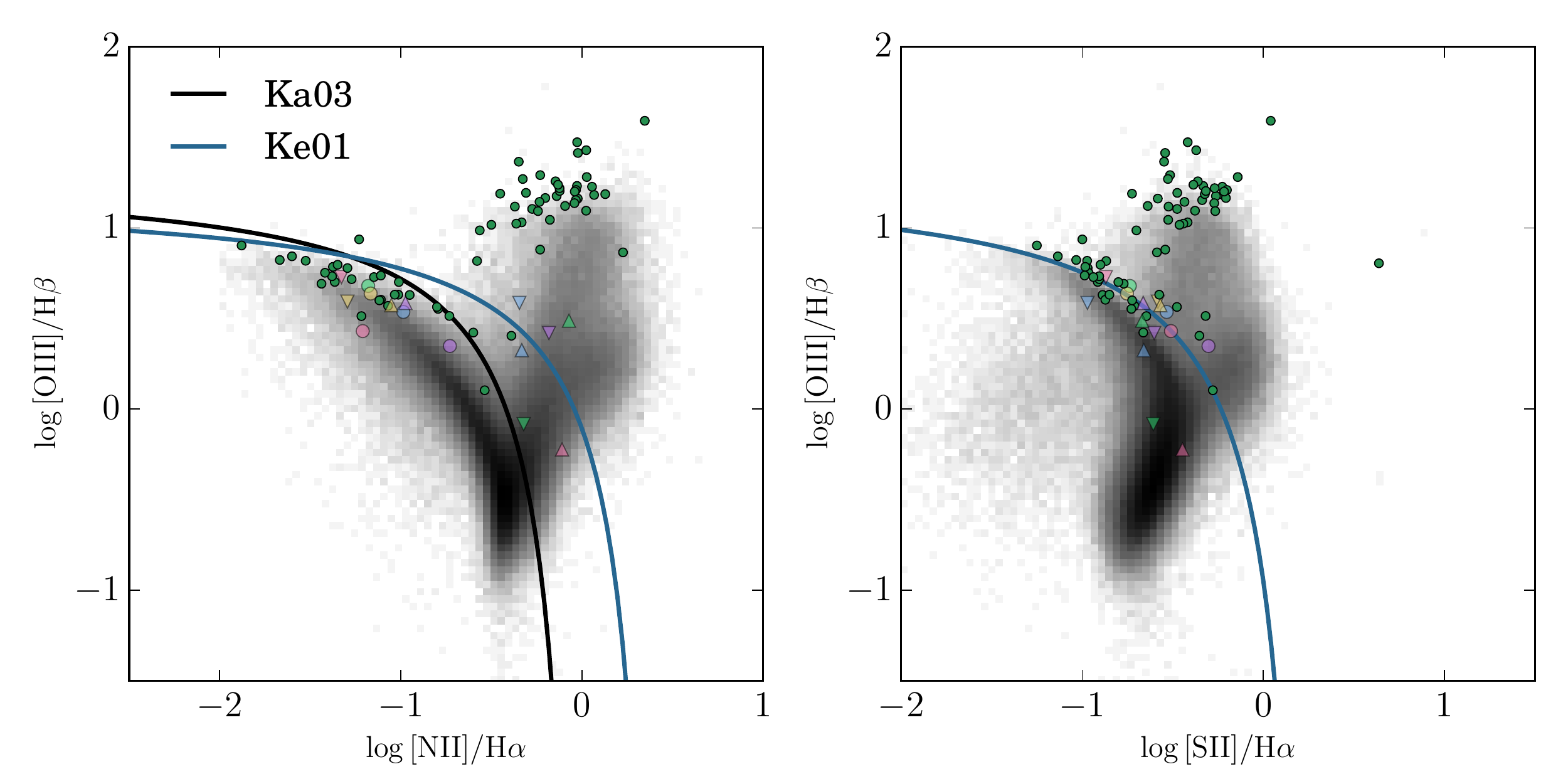}
\caption{
BPT diagrams based on the $\mathrm{[NII]/H\alpha}$ and $\mathrm{[OIII]/H\beta}$ emission line ratios (left panel) and $\mathrm{[SII]/H\alpha}$ and $\mathrm{[OIII]/H\beta}$ ratios (right panel). The grey background represents the density of SDSS galaxies (927\,552 galaxies from DR7), on a logarithmic colour scale. Outlying galaxies are marked with green circles. Outlier galaxies which are classified as HII regions in one diagnostic and as AGN in another are each plotted with the same marker and colour in both panels. The classification limits are taken from the extreme starburst limit by \citet{kewley01} (Ke01; blue line) and composite galaxies by \citet{kauffmann03a} (Ka03; black line).
}\label{f:bpt_all}
\end{figure*}

We find two additional galaxies with coronal lines ([FeX] 6376\AA, [FeVII] 6088\AA, [FeVII] 5722\AA, [FeXIV] 5304\AA, [FeVII] 3759\AA), these galaxies do not meet the criteria of \citet{wang12}, since they are classified as AGN in the BPT diagnostic diagram. We show an example of such galaxy, SDSS J120041.39+314746.2 (6473-56363-672), in figure \ref{f:CLE_example_plot} where one can see the emission lines [FeVII] 6088\AA, [FeVII] 5722\AA, and [FeXIV] 5304\AA. \citet{gelbord09} look for forbidden high ionisation lines in the SDSS DR6 and find 63 objects, some of which are classified by the SDSS pipeline as galaxies and some as QSOs. 

\subsubsection{Visible $\mathrm{[NI]}$ emission line}
We find ten galaxies which show an emission line at 5200\,\AA, for eight of them one must subtract the best-fitting population synthesis model in order to detect the emission line (as done is Section 4.1.1). We show an example of such galaxy in figure \ref{f:ni_single_plot}. SDSS J102339.63+041110.6 (575-52319-200) is a galaxy at redshift $z=0.290$, classified as a composite (i.e., both HII region and AGN radiation are prominent radiation sources) using the BPT diagram. Furthermore, the galaxy is a member of a galaxy cluster and is detected in radio emission (FIRST catalog).

We fit Gaussians to the emission lines $\mathrm{H\beta}$, $\mathrm{[OIII]}$, $\mathrm{[NII]}$, $\mathrm{H\alpha}$, $\mathrm{[SII]}$, and the line at 5200\,\AA. We find that the FWHM of the 5200\,\AA\, emission line is consistent with the FWHM of the forbidden lines for both of the galaxies. We identify the emission line at 5200\AA\, as emission from $\mathrm{[NI]}$, using the emission-line catalogue conducted by \citet[see references therein]{kaler97}. $\mathrm{[NI]}$ emission is usually observed in planetary nebulae and its strength is typically 10-100 times weaker than $\mathrm{H\beta}$ emission, it is also detected in composite quasar spectra with a typical strength that is 10 times weaker than $\mathrm{H\beta}$ \citep{zakamska14}. In our galaxies the EW ratio of $\mathrm{[NI]/H\beta}$ ranges from 0.33 to 1. However, we note one case in the literature where the $\mathrm{[NI]}$ emission strength is comparable to the $\mathrm{H\beta}$ emission, PKS 2322-12 (\citealt{costero77}), which is a narrow-line radio galaxy.

While galaxies often show strong emission from singly-ionised nitrogen ($\mathrm{[NII]}$), atomic nitrogen is not observed in emission since it is too easy to ionise. Atomic nitrogen can exist in cool areas where it is shielded from ionising radiation and can be seen in weak emission, most likely due to photo-excitation. An alternative scenario, which can potentially produce stronger $\mathrm{[NI]}$ emission, can take place in cool cluster filaments, where the shielded atomic nitrogen is excited by shocks or hard penetrating X-ray photons \citep{ferland09}. These scenarios however do not explain how we see such strong emission in $\mathrm{[NI]}$ in these galaxies. This may require photoionisation calculations that are beyond the scope of this paper. 

\subsection{Unusual emission lines ratios}\label{s:tails}
Most of the galaxies in this subsample are of known classes, but their emission lines lie at the extreme of their population's respective distribution. These objects show outlying line ratios -- 156 of them are outliers on the BPT line diagnostic diagram, and there are 30 galaxies for which only the $\mathrm{H\alpha-[NII]-[SII]}$ complex is detected. Furthermore, they show relatively weak $\mathrm{H\alpha}$ emission compared to $\mathrm{[SII]}$ and $\mathrm{[NII]}$ emission. 

\subsubsection{Outliers on the BPT diagram}\label{ss:emission_bpt}
Prominent emission lines in galaxies have been used to study the physical properties of galaxies and their dominant source of radiation. \citet{baldwin81} proposed a diagnostic diagram (known today as the BPT diagram) based on the ratios $\mathrm{[NII]/H\alpha}$ and $\mathrm{[OIII]/H\beta}$ to separate between gas that is exposed to hard radiation fields, planetary nebulae, and HII regions. \citet{veilleux87} refined this scheme and added two additional diagnostic diagrams with the ratios $\mathrm{[SII]/H\alpha}$ and $\mathrm{[OI]/H\alpha}$, based on photoionisation models. The SDSS, providing millions of galaxy spectra, has had a great impact on the classification of galaxies and allowed a clean and significant characterisation of the different types of galaxies and the transitions between them \citep{kauffmann03b, kewley06}.

\begin{figure*}
\includegraphics[width=0.45\textwidth]{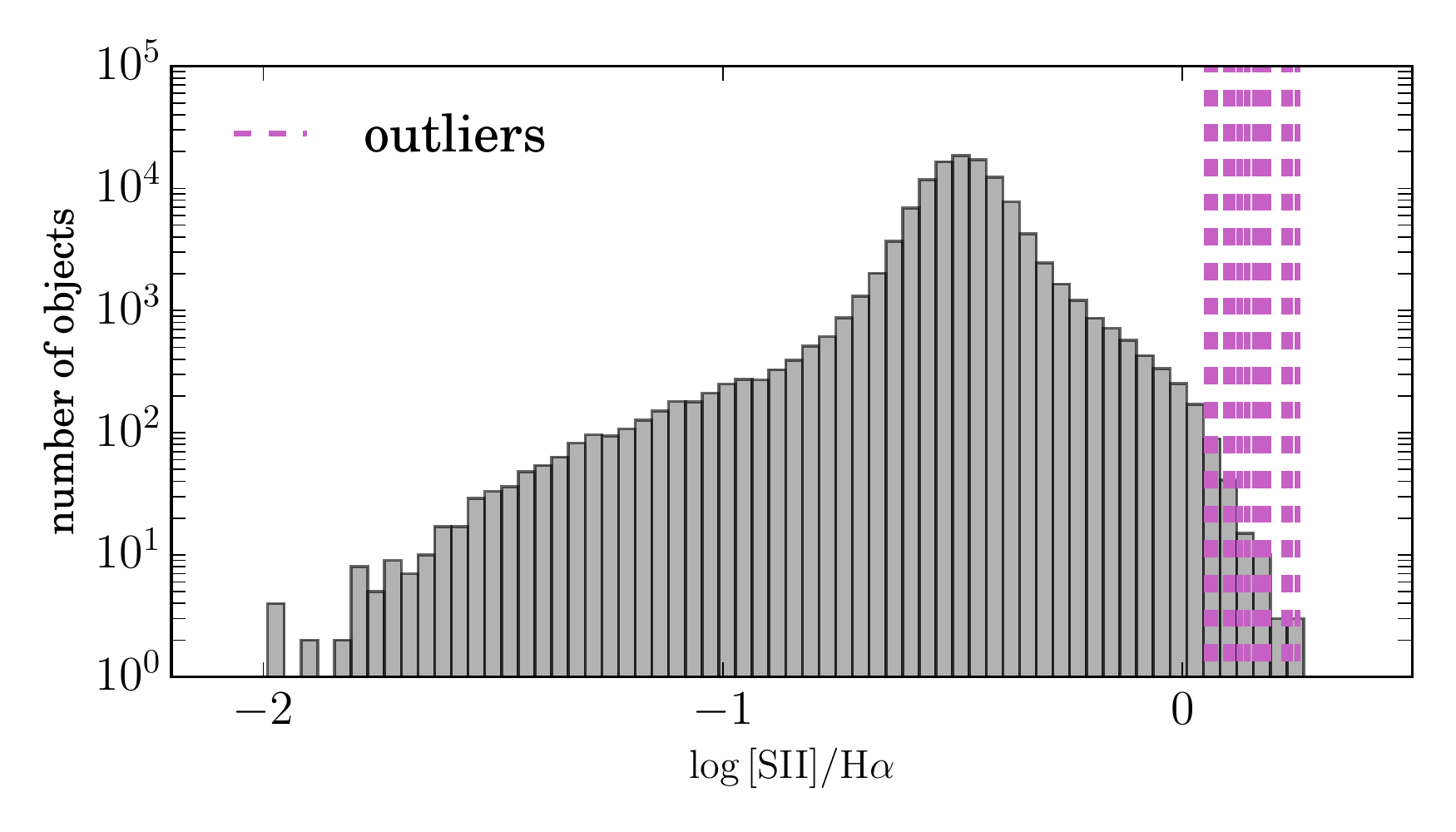}
\includegraphics[width=0.45\textwidth]{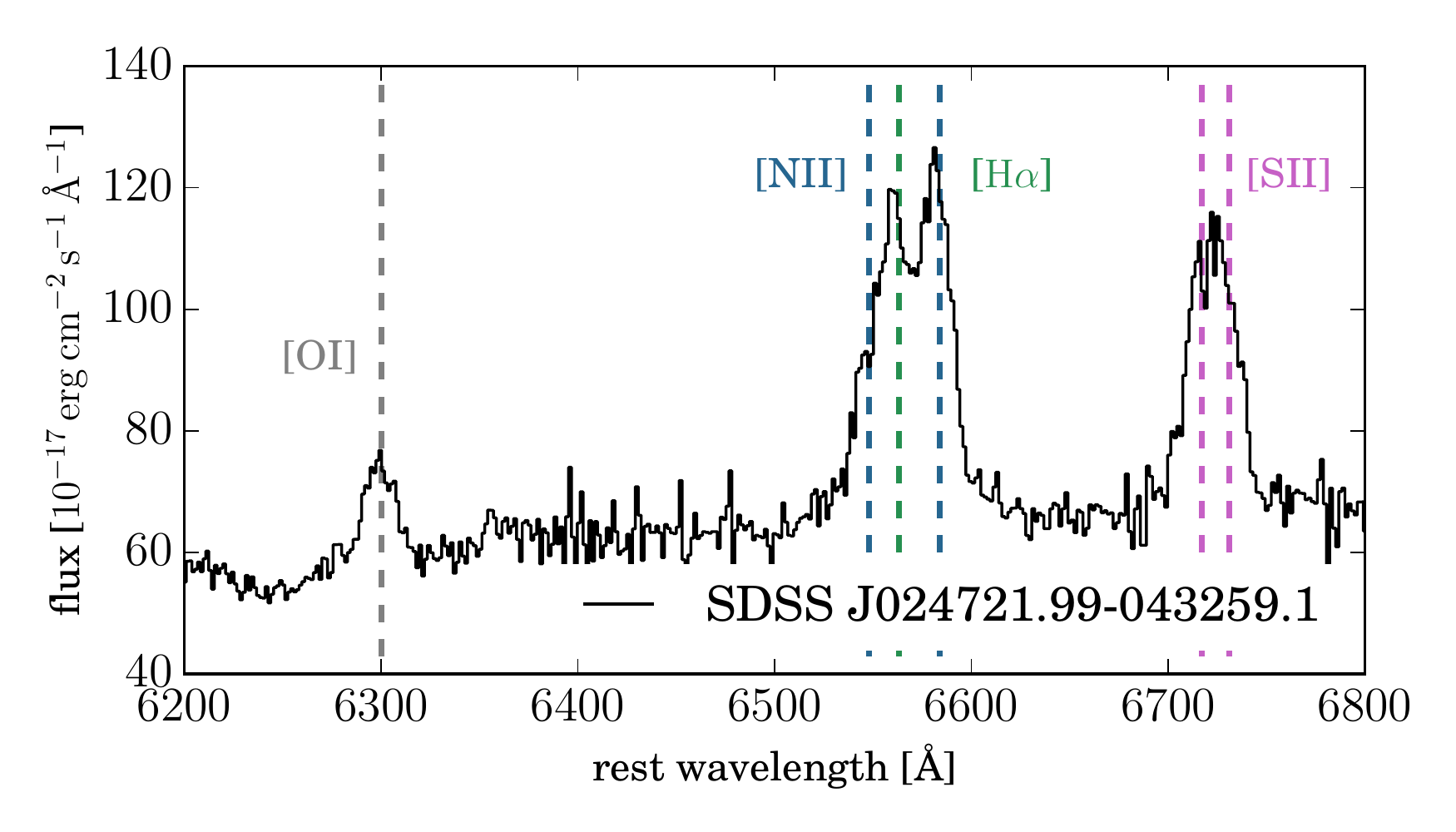}
\caption{Comparison between the $\mathrm{log\,[SII]/H\alpha}$ line ratio of the 24 outlying galaxies (pink) and the general galaxy population (grey histogram) using the MPA-JHU catalogue of SDSS galaxies (left panel). One can see that the galaxies we find are located on the tail of the general distribution and have unusual line ratios, greater than most of the galaxies in SDSS. We give an example of such a galaxy (SDSS J024721.99-043259.1; 7054-56575-500), at redshift $z=0.138$ which has a ratio $\mathrm{log\,[SII]/H\alpha} = 0.17$.
}\label{f:s_halpha_hist}
\end{figure*}

We present in figure \ref{f:bpt_all} a BPT diagram for the emission line ratios: $\mathrm{[NII]/H\alpha}$, $\mathrm{[SII]/H\alpha}$, and $\mathrm{[OIII]/H\beta}$. We
show typical galaxies from the MPA-JHU catalogue in grey, on a logarithmic colour scale. For our outlying galaxies, we measure the emission lines $\mathrm{[SII]}$, $\mathrm{[NII]}$, $\mathrm{H\alpha}$, $\mathrm{H\beta}$, and $\mathrm{[OIII]}$ by integration over Gaussians profiles that we fit: the $\mathrm{H\alpha}$, $\mathrm{[NII]}$ and $\mathrm{[SII]}$ are modelled jointly with 5 Gaussians at fixed rest-frame wavelengths and the $\mathrm{[OIII]}$ and $\mathrm{H\beta}$ profiles are similarly measured with 3 Gaussians. We examine the goodness of the fit by-eye, and when needed we add additional Gaussian component (e.g., if a profile is not symmetric) or allow the rest-frame wavelength to shift. We mark the emission line ratios of our outliers in green. 

\citet{kewley05} analysed the effect of a fixed aperture size on metallicity, star formation rate, and reddening and concluded that a minimum aperture covering a fraction of $\approx 20$ precent is required for the spectral properties to approximate the global values, which translates to a lower limit on redshift of $z > 0.04$ (for the SDSS spectrograph). Out of the 156 emission line galaxies with unusual ratios, only 86 satisfy this criterion. One can see that the outlier galaxies show emission line ratios that reside on the tails of the distribution of the entire population, which explains their weirdness score.

15 outliers are classified as HII regions with one diagnostic and AGN with another diagnostic. We show in figure \ref{f:bpt_all} examples for such cases. We use two separating criteria: \citet{kewley01} who used a combination of population synthesis models and photoionisation models to produce a theoretical upper limit that separates between starbursts and AGN-dominated galaxies (Ke01; blue line), \citet{kauffmann03a} modified this limit by including composite galaxies which contain significant contribution from both star formation and AGN (Ka03; black line). Individual objects that show this duality are marked with the same colour and symbol in the two panels, for convenience. We list the 156 BPT outliers in table \ref{table:bpt}.

\subsubsection{Weak $\mathrm{H\alpha}$ emission}\label{ss:emission_s}
We identify 30 galaxies with emission in $\mathrm{H\alpha}$, $\mathrm{[NII]}$, and $\mathrm{[SII]}$ but none in $\mathrm{H\beta}$ and $\mathrm{[OIII]}$. Therefore, in order to study their gas emission properties, we fit and subtract a stellar population synthesis model obtained (as in section 4.1.1). These galaxies have unusually large $\mathrm{[NII]}$ and $\mathrm{S[II]}$ over $\mathrm{H\alpha}$ ratios ($\mathrm{log\,[SII]/H\alpha} \approx 0.1$). Some of the objects are low redshift galaxies, with an angular size a few times the diameter of the SDSS fibre, so the spectrum only samples a small fraction of the gas near the galactic centre.

However, we find 7 objects with $z > 0.065$ and by comparison of their Petrosian radius (from SDSS photometry) to the fibre diameter we verify that the SDSS fibre indeed captures 20 percent or more of the galactic light. We measure the $\mathrm{log\,[SII]/H\alpha}$ line ratio for our objects and present it in figure \ref{f:s_halpha_hist} compared to the line ratio histogram of the MPA-JHU catalogue of galaxies from the SDSS (left panel), where one can see that our objects reside on the tail of this distribution. We give an example of such a galaxy at redshift $z=0.138$ for which $\mathrm{log\,[SII]/H\alpha} = 0.17$ (right panel). We list these galaxies in table \ref{table:snh}.

High $\mathrm{log\,[SII]/H\alpha}$ and $\mathrm{log\,[NII]/H\alpha}$ emission line ratios are generally attributed to LINER galaxies (see for example \citealt{kewley06} and references within), though the ratios we measure for our outlier galaxies deviate from the typical ratio produced by a LINER galaxy. These galaxies can therefore be extreme cases of LINER galaxies in which we do not detect the $\mathrm{H\beta}$ and $\mathrm{[OIII]}$ emission due to limited SNR, or high dust extinction along the line of sight to the AGN. Alternatively, the high emission line ratio can be due to shock ionisation from galactic-scale winds \citep{heckman80, dopita95, lipari04, veilleux05, allen08}. One can find in the literature additional examples of high $\mathrm{log\,[SII]/H\alpha}$ emission line ratios measured with integral field spectroscopy within individual lines of sight in galaxies (e.g. \citealt{singh13} and \citealt{cheung16}).

\begin{figure*}
\includegraphics[width=0.34\textwidth]{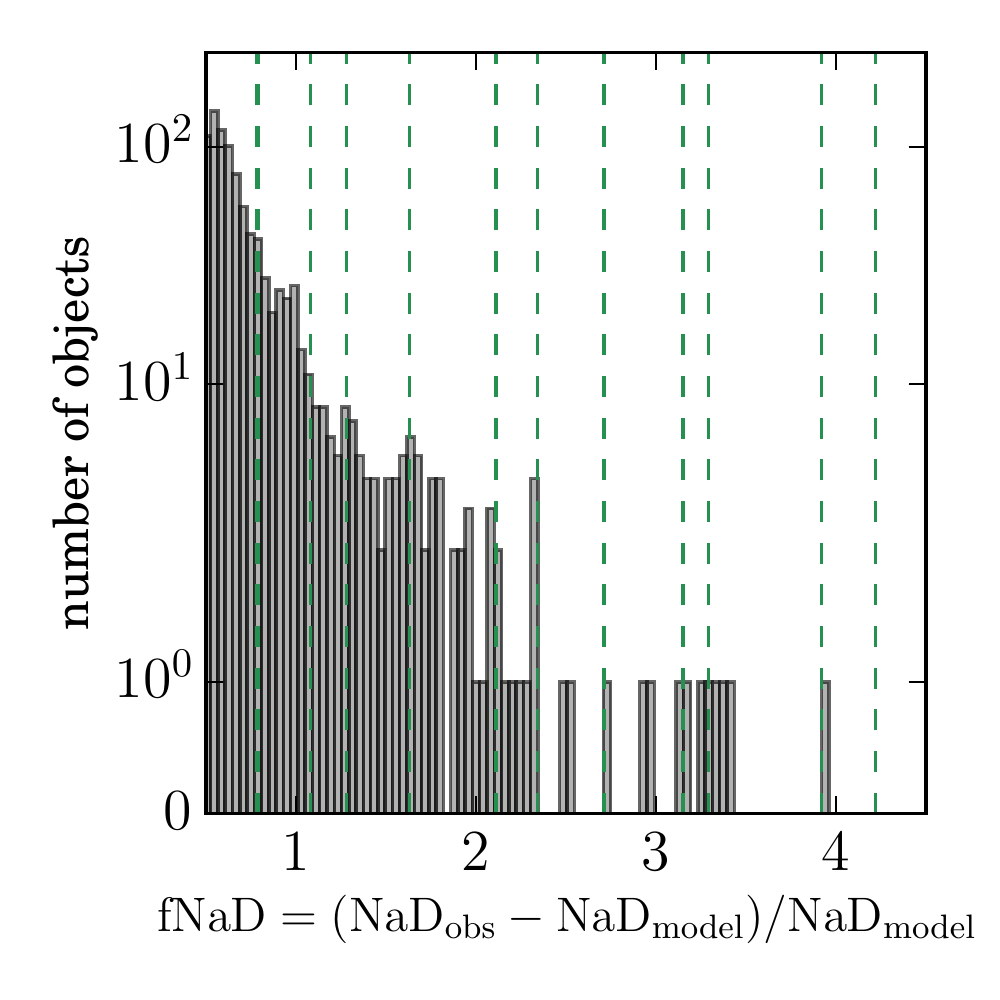}
\includegraphics[width=0.60\textwidth]{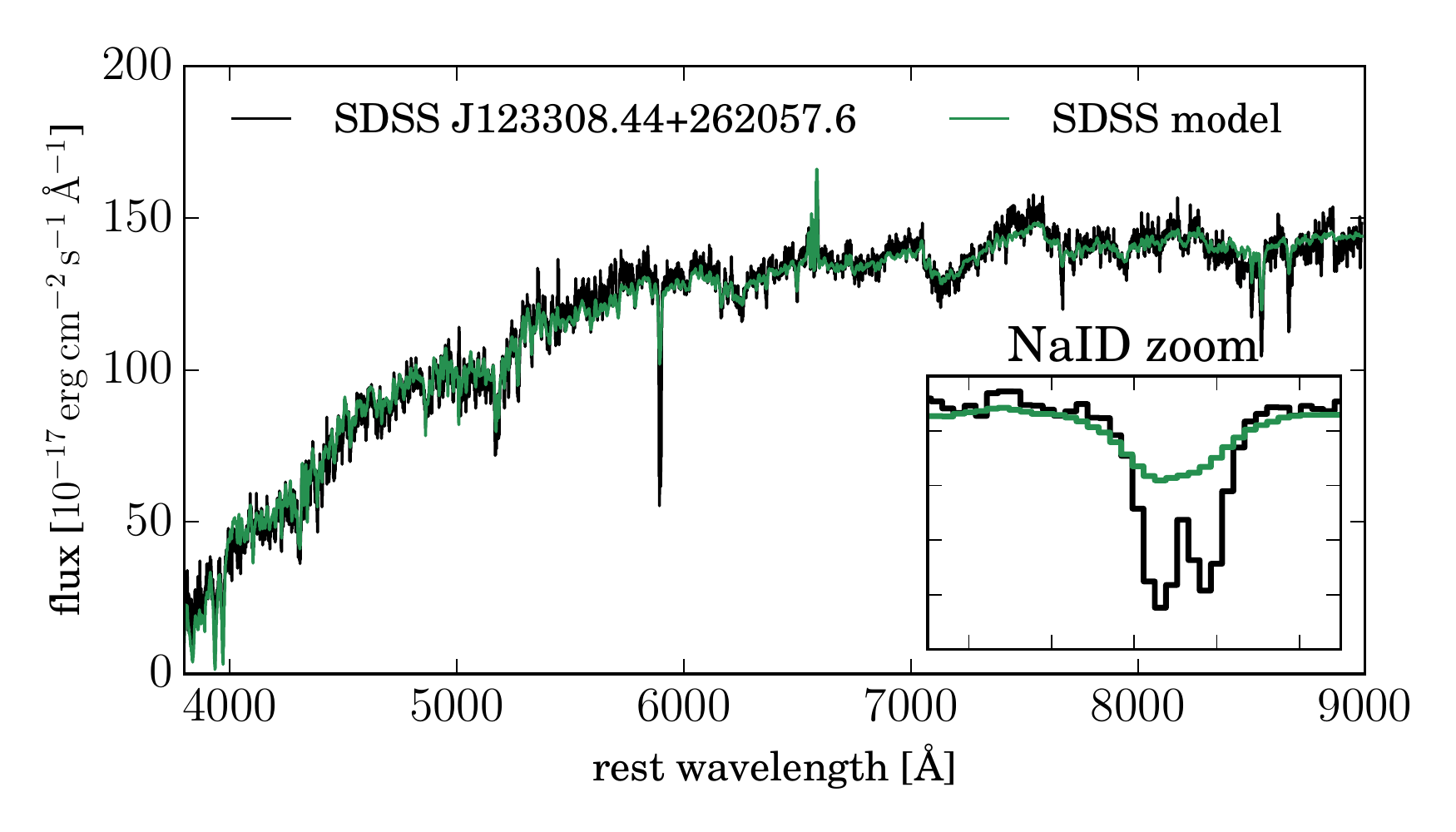}
\caption{Histogram of NaID excess index (see section \ref{ss:sod} for definition) of the NaID excess objects studied by JE13 (grey bars), these are the 1603 galaxies which have the most extreme discrepancy between the observed NaID and the NaID predicted from a standard model of a galaxy. We mark the indices of the 13 outlying objects our algorithms finds, which are also present in the JE13 sample. The right panel presents the galaxy SDSS J123308.44+262057.6 (2235-53847-0086), which is not included in the JE13 catalogue, at redshift $z = 0.024$ (black), compared to the SDSS best fitting model (green). One can see that the observed NaID absorption is more than double what is expected from to the best fitting model.
}\label{f:sod_axample}
\end{figure*}

\subsection{Extremely red and sodium excess galaxies}\label{s:red_and_sod}
We find 47 galaxies that have extremely red continua and/or unusually strong sodium absorption. 

\subsubsection{Sodium excess galaxies}\label{ss:sod}
38 galaxies have very strong sodium (NaID 5889, 5895\AA) absorption, some with extremely red continua as well. There is some debate in the literature regarding the origin of the NaID excess objects -- one explanation is that the NaID excess is related to the interstellar medium (ISM), either due to galactic-scale gaseous outflows in star forming galaxies or cool gas that resides in the disk, another explanation is that the absorption is stellar, and due to variations in metal abundance. \citet[and references therein]{vandokkum10} suggested that an initial mass function (IMF) in early-type galaxies with a larger number of low-mass stars (less than 0.3 M$_{\odot}$) may lead to stronger NaID absorption.

In order to address these questions, \citet[hereafter JE13]{jeong13} compiled a catalogue of roughly 1000 NaID excess objects up to SDSS DR7. They limited themselves to the redshift range $z\le 0.08$ and applied further colour cuts to ensure the completeness of their sample, from which they identified the NaID excess objects. The authors used the absorption line measurements of all the galaxies in the SDSS performed by \citet{oh11} and defined the index $fNaD$ as follows:
\begin{equation}\label{eq:2}
	{fNaD = \frac{NaD_{observed} - NaD_{model}}{NaD_{model}}}
\end{equation}
where $NaD_{observed}$ is the observed Na D line strength and $NaD_{model}$ is the expected model Na D line strength. The expected $NaD_{model}$ is also taken from \citet{oh11}, where the authors fitted population synthesis models from \citet{bruzual03} and stellar templates by \citet{sanchez06}), and calculated
internal reddening due to dust. JE13 chose 1603 galaxies for which $fNaD > 0.5$ (i.e., observed strength is 50\% stronger than expected by model; see figure 1 in JE13) and defined them as their sodium excess objects. 

We find 24 galaxies that were not part of the JE13 sample, some for obvious reasons such as the redshift cut, the SDSS data release, and the colour cuts they performed. We find 4 galaxies which satisfy the JE13 cuts in DR7 but were not part of their sample, possibly due to erroneous measurement of the observed Na D. We present in the left panel of figure \ref{f:sod_axample} the histogram of the NaID excess objects from JE13 compared to the outlying galaxies we find. In the right panel of figure \ref{f:sod_axample} we show an example of such outlying galaxy, at redshift $z = 0.024$, compared to the best SDSS model of this galaxy. One can see that the observed NaID is more than double the strength expected from the model. We list the NaID excess objects in table \ref{table:sod}.

\subsubsection{Extremely red galaxies}\label{ss:red}
We find 19 galaxies which are extremely red, most of them show a diagonally-shaped continuum. We use the galaxy composite catalogue by \citet{dobos12} (hereafter DO12). DO12 classified all the galaxies in SDSS DR7 by colour, nuclear activity, and star formation activity and constructed a set of spectra with high SNR and high resolution. They fitted the continuum of the composite spectra using the population synthesis models by \citet{bruzual03} and found that the composite spectra represent the different galaxy populations well. Using the best stelar population synthesis model, they extracted the best metallically and the optical depth of interstellar dust. The DO12 composites are divided into colours -- red, green, and blue -- and into classes -- passive, star forming, $\mathrm{H\alpha}$ detected, AGN and $\mathrm{H\alpha}$ detected (for objects which show both nuclear activity and substantial star formation), LINERs, and Seyfert galaxies. 

\begin{figure*}
\includegraphics[width=1\textwidth]{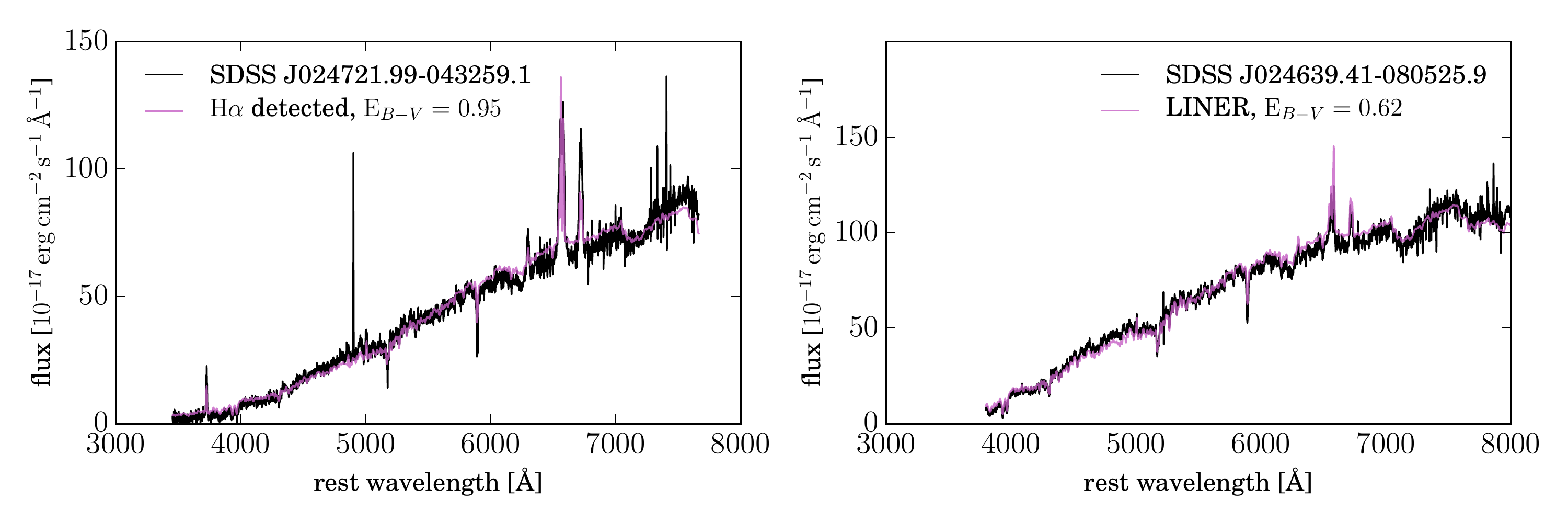}
\caption{Examples of two galaxies which show diagonally-shaped continua. We fit the galaxies with the galaxy composites constructed by DO12 with additional reddening in order to match their red colours. SDSS J024721.99-043259.1 (7054-56575-500; left panel; black) is best reproduced with a `$\mathrm{H\alpha}$ detected' composite with total dust reddening of $\mathrm{E}_{B-V} = 1.1$ (pink) and SDSS J024639.41-080525.9 (7057-56593-115; right panel; black) is best reproduced with a LINER composite with total $\mathrm{E}_{B-V} = 0.98$. 
}\label{f:dust_red}
\end{figure*}

Composite spectra which belong to the same class of galaxies (passive for example) but in a different colour bin differ not only in dust optical depth but also in metallicity and stellar population. We therefore use all the DO12 composite spectra and apply additional dust reddening to them in order to reach the continuum shape we observe in our galaxies. We use the prescription by \citet{fitzpatrick99} for the additional dust extinction and find the best composite spectrum and the additional dust reddening to be applied to this composite in order to reproduce the continuum shape of our galaxies. We find that for all of our objects the reddest composite spectrum is a better fit than the green and the blue. For these DO12 derive metallicity value of $Z=0.02$.

The dust reddening we find ranges from $\mathrm{E}_{B-V} = 0.3$ to $\mathrm{E}_{B-V} = 0.9$ for our outlying galaxies, this is in addition to the dust that is derived by DO12 in their stellar population synthesis models, giving in total colour excesses between $\mathrm{E}_{B-V}$ = $0.5$ and $2.2$. We show two examples of extremely red galaxies and their best DO12 composite plus additional reddening in figure \ref{f:dust_red}. We find that the best fitting composite for SDSS J024721.99-043259.1 (7054-56575-500; left panel) belongs to the $\mathrm{H\alpha}$ detected class, with colour excess derived from population synthesis of $\mathrm{E}_{B-V} = 0.15$ but with an additional $\mathrm{E}_{B-V} = 0.95$. The second object (right panel), SDSS J024639.41-080525.9 (7057-56593-0115), is best described by a LINER galaxy composite with a total of $\mathrm{E}_{B-V}=0.98$. We list these galaxies in table \ref{table:red}, where we note specifically the galaxy SDSS J044023.25+245402.0 (1257-52944-314) with $\mathrm{E}_{B-V}=1.6$, caused by MW dust. These galaxies are rare in SDSS due to colour selection.

\subsection{Galaxies hosting supernovae}\label{ss:sne}
We find 18 galaxies in which a supernova happens to have exploded when the spectrum was taken. Supernova rates are of order one per galaxy per century (e.g., \citealt{li11}), or one per 5000 weeks. Assuming a supernova maintains a luminosity comparable to its host's for about a week, there should be hundreds of such objects in the SDSS galaxy sample, which is rare but not ridiculously so. There are several studies whose main goal is the detection of supernovae in the galaxy spectra of the SDSS, and they found 15 of our objects. In all of these studies, the galaxy spectra are fitted with galaxy eigenspectra that are obtained from PCA analysis of SDSS galaxies. supernovae detection is then carried out on the residual between the galaxy spectrum and its best eigenspectra fit, using various methods of template fitting: \citet{madgwick03a} used wavelet transform in order to obtain a residual which is noiseless, and then cross-correlated it with a set of type Ia spectral templates. \citet{tu10, krughoff11, graur13, graur15} used PCA or singular value decomposition (SVD) of supernovae spectroscopic templates to fit the residual. In all of these cases, it was necessary to use both galaxy and supernovae templates in order to detect the supernovae. We list the outlying galaxies with supernovae in table \ref{table:sne}.

\begin{figure*}
\includegraphics[width=1\textwidth]{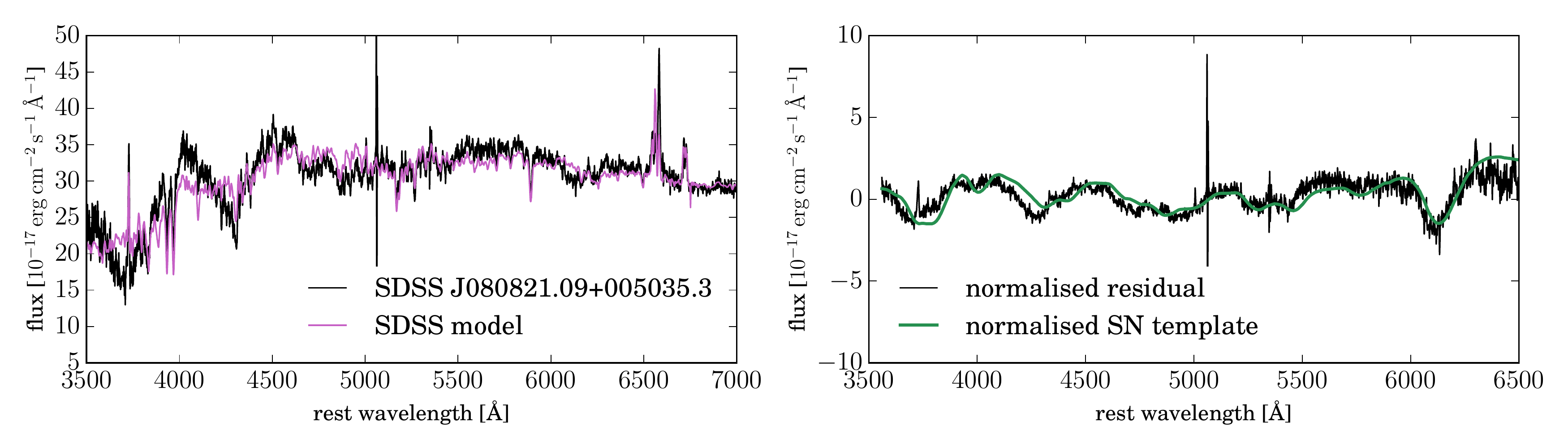}
\caption{SDSS J080821.09+005035.3 (4745-55892-0018) spectrum and its best SDSS model (left panel), where one can see that the SDSS model fails to account for some broad features in the spectrum. We subtract the model from the spectrum and normalise it with a continuum fit (right panel; black) and compare it to a supernova template (2 days after the peak) which we also normalise (green). One can see that the supernova template is able to account for the residual broad features in the galaxy spectrum.
}\label{f:SN_fit}
\end{figure*}

SDSS J080821.09+005035.3 (4745-55892-0018) is reported here for the first time since the galaxy is part of DR10, while \citet{graur15} worked with the DR9 galaxy sample. We present in figure \ref{f:SN_fit} the galaxy spectrum and its best SDSS model (left panel) where one can notice additional, broad, features that the best SDSS model does not account for. We subtract the best fitting SDSS model from the spectrum and normalise the residual with a continuum fit and compare the residual to a type Ia supernova template from \citet{nugent02} 2 days after $B$-band maximum. This is shown in the right panel of figure \ref{f:SN_fit}. One can see that the supernova template matches well the broad residual features in our galaxy.

SDSS J085425.56+180507.0 (2281-53711-0149) and SDSS J085436.69+180552.9 (2281-53711-0156) are also reported here for the first time. These spectra were in DR7, but the SDSS pipeline failed to fit the correct redshift to them, mistaking the $\mathrm{H\alpha}$ emission for $\mathrm{Ly\alpha}$ at high redshift. We derive the correct redshift by fitting the narrow $\mathrm{[OIII]}$ and $\mathrm{H\alpha}$ emission lines, finding redshifts of 0.094 and 0.211 respectively. For this reason they were missed by \citet{graur15}. Using the same fitting procedure as described in \citet{graur13} and \citet{graur15}, we find that SDSS J085425.56+180507.0 is either a type Ia or a type Ic near maximum light. SDSS J085436.69+180552.9 is most likely a type Ia near peak (O. Graur, private communication).

\subsection{Blends}\label{s:blends}
We find 27 outliers which are blends of two objects. These include 6 galaxy-galaxy gravitational lenses, 9 spectra with two sets of emission lines, and 12 spectra which appear unusual due to chance alignment of a galaxy and a nearby star. We list the galaxies which are contaminated by nearby stars in table \ref{table:chance_al}.

Multiple studies were carried out with the goal of finding gravitationally lensed galaxies in the SDSS spectroscopic data \citep{bolton04, bolton05, bolton06, bolton08, brownstein12, tsalmantza12, shu16}. These studies focused on galaxies with two sets of nebular emission lines at redshifts that differ significantly. For example, \citet{bolton04} used the best fitting model obtained by the SDSS pipeline and subtracted it from the spectrum, resulting in a residual spectrum. The residual spectrum underwent a match-filtering algorithm that was designed to find an additional set of emission lines. \citet{tsalmantza12} used a data-driven model which recovered most of the known galaxy lenses in the SDSS by constructing basis functions for dimensionality reduction. Overall, up to DR12, these studies have found 392 galaxy lens candidates in the SDSS. We list the six galaxy lenses our algorithm recovers in table \ref{table:g_lens}, all of which were reported by previous studies. 

\begin{figure*}
\includegraphics[width=0.35\textwidth]{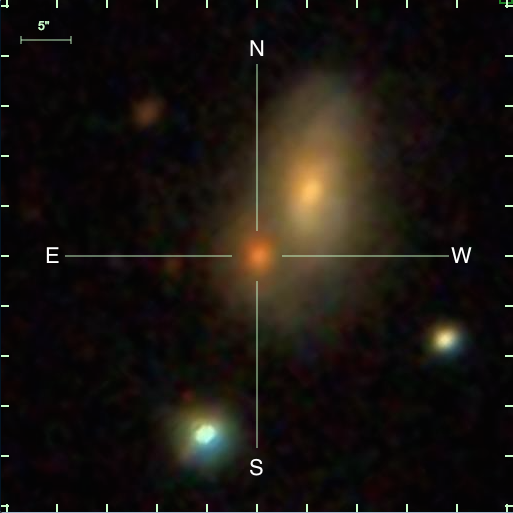}
\includegraphics[width=0.55\textwidth]{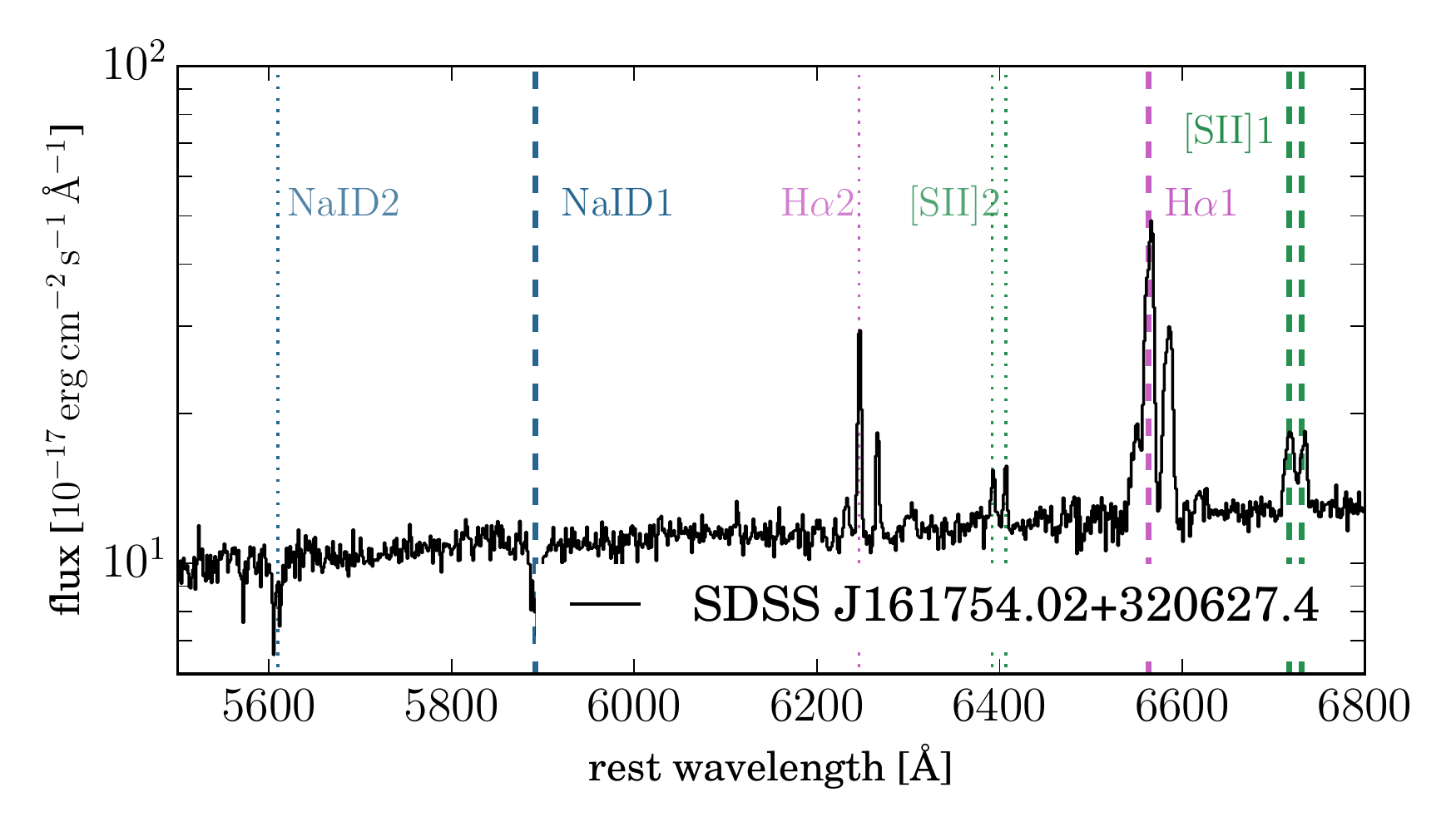}
\caption{Example of a galaxy with two sets of emission lines. The primary galaxy (SDSS J161754.02+320627.4; 1420-53146-528), marked with a cross in the image from the SDSS (left panel), is at redshift $z=0.117$. The secondary galaxy (SDSS J161753.62+320633.8) is at redshift $z=0.063$ and the gas in its outskirts is along the line of sight towards the primary galaxy. The spectrum of the primary galaxy (right panel) consists of two sets of emission lines, one at redshift $z=0.117$ and the second at redshift $z=0.063$.}\label{f:multilines_examp}
\end{figure*}

The second subset consists of objects with two sets of emission lines that were not previously reported as galaxy lenses. The second set of nebular emission lines is due to a close companion (i.e., at a small angular separation) to the primary galaxy, and in most of the objects the primary and the companion galaxy are  somewhat to well distinguished in the SDSS imaging. For each of the objects, we fit two sets of emission lines with Gaussians in order to ascertain that all the nebular lines of the secondary set differ by the same velocity compared to the primary set of lines and measure the redshift of the secondary object. In all of the cases, the redshift produced by the SDSS pipeline belongs to the more luminous set of emission lines. We find that for 7 out of 8 of the objects the secondary set of emission lines is emitted by gas at a lower redshift than the primary. For three of the pairs we find an SDSS spectrum of the secondary galaxy as well as the primary, and find that the redshift estimation of the secondary galaxy is equal to the redshift we derive from the second set of emission lines in the primary's object spectrum. We present an example of such galaxy in figure \ref{f:multilines_examp}, where we show the SDSS image of the primary and secondary galaxy (left panel; the cross indicates the primary object) and the spectrum of the primary, shifted to its rest-frame, where one can see two sets of emission lines. The first set of emission lines is at $z=0.117$ and the second set at $z=0.063$, which is the redshift of the secondary galaxy.

We list all the objects in table \ref{table:multi_lines}, and note the redshift of the primary and secondary sets of lines. One of the objects, SDSS J160818.74-002745.4 (345-51690-229), is an exception since no clear companion can be seen in the imaging. Furthermore, the secondary set of emission lines is at a higher redshift than the primary galaxy, and these two sets are separated only by $2860\,\mathrm{km\,s^{-1}}$. The spectrum provided by the SDSS is a median spectrum which consists of a few sub-exposures of the same galaxy, each of them undergoes a separate wavelength solution, sky subtraction, and flux calibration. The calibrated spectra are then combined using a median filter. A wrong wavelength solution could cause the median spectrum (with which we work) to appear as a spectrum with two (or more) sets of emission lines. We compare the sub-exposures of J160818.74-002745.4 and find two sets of emission lines in all of them, indicating that this is not an instrumental error.

\begin{figure*}
\includegraphics[width=1\textwidth]{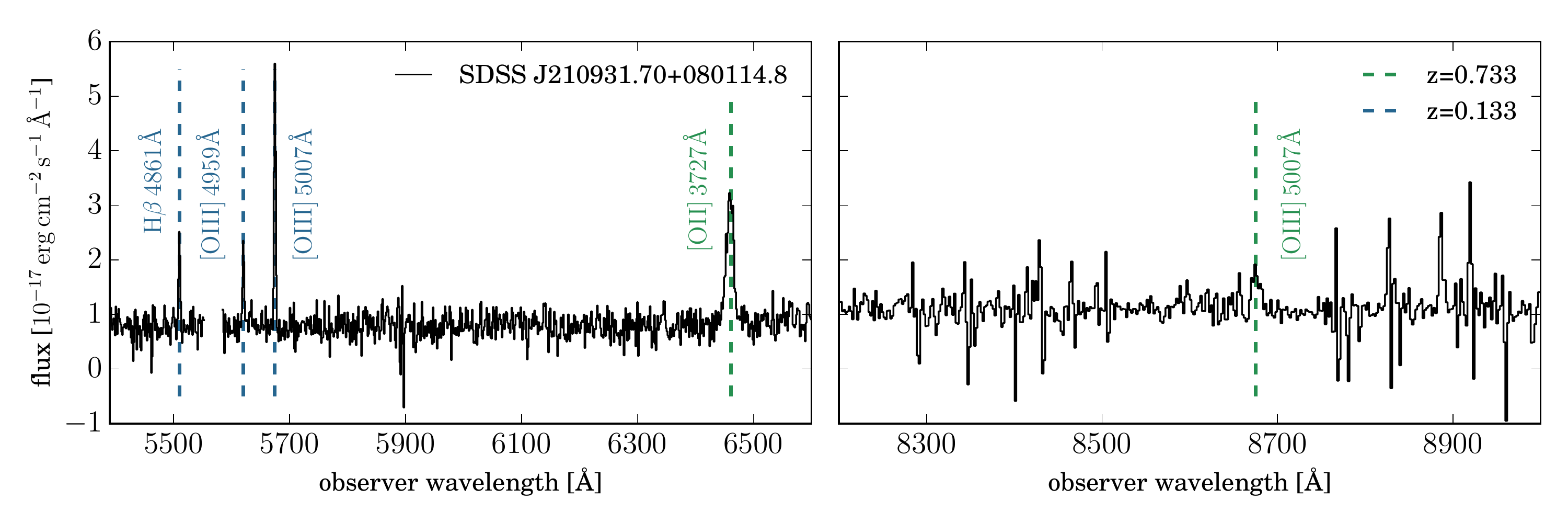}
\caption{SDSS J210931.70+080114.8 (4079-55363-773) is a galaxy at redshift $z=0.133$ for which we detect an emission line in wavelength 6460\,\AA\, in the observers frame. This emission line is broader than the emission lines that belong to this galaxy (left panel) and does not match in wavelength to any known emission lines at redshift $z=0.133$. Assuming the emission line is due to $\mathrm{[OII]}$ 3727\,\AA\, in the rest-frame of a background galaxy, we estimate the redshift of the background galaxy to be $z=0.733$. We mark in the right panel the wavelength of $\mathrm{[OIII]}$ 5007\,\AA\, in the observers frame for the redshift $z=0.733$, which matches a weak feature in the spectrum. Since the flux of the supposed $\mathrm{[OII]}$ emission is within the range expected for galaxies at redshift $z=0.7$, we conclude that this galaxy is another case of galaxy-galaxy chance alignment.
}\label{f:blend_candidate}
\end{figure*}

We note an additional object, SDSS J210931.70+080114.8 (4079-55363-773), for which the secondary redshift estimation is based only on two emission lines, one of which is marginally detected. The primary redshift of the galaxy is $z=0.133$ and no clear companion is detected in the imaging. However, we find an unidentified emission line in wavelength 6460\,\AA\, in the observer’s frame, and its FWHM is inconsistent with the FWHM of the emission lines of the galaxy. Assuming the unidentified emission line is due to $\mathrm{[OII]}$ 3727\,\AA\, in the rest-frame of a background galaxy, we estimate the redshift of the background galaxy to be $z=0.733$. We show the spectrum of the galaxy in figure \ref{f:blend_candidate}, where the supposedly $\mathrm{[OII]}$ emission is marked with a green dashed line (left panel). In the right panel of figure \ref{f:blend_candidate} we mark the wavelength of $\mathrm{[OIII]}$ 5007\,\AA\, in the observers frame for the redshift $z=0.733$, which match a weak feature in the spectrum. The FWHM values of both of them are consistent with each other. The flux of the $\mathrm{[OII]}$ emission is in the range expected from star forming galaxies at redshift $z=0.733$, thus no magnification is needed in order to explain the $\mathrm{[OII]}$ emission strength. We therefore conclude that their background galaxy is not strongly lensed.

\subsection{Calibration errors, source confusion and sky contamination}\label{s:errs}
This group of outliers is a mixed bag of objects that are broadly defined as not being astronomically `interesting'. This includes calibration errors and SDSS pipeline misclassifications. In section \ref{s:data} we remove pixels that are marked as bad by the SDSS pipeline, and we also apply a median filter for the removal of residual cosmic rays and unresolved sky emission. The SDSS also provide quality flags for all the plates and all their spectra. Since we do not limit ourselves to plates of good quality only, it is inevitable that some bad spectra exist in our dataset. Additionally, our algorithm finds the relatively few objects that were misclassified by the SDSS pipeline, and we include them here. 

First, we identify 41 galaxy spectra that are in fact single stars. Additionally, we find 5 spectra of white dwarfs that were mistakenly identified as galaxies, 4 of these were previously reported as white dwarfs by \citet{girven11} and \citet{kleinman13}, and one of them, SDSS J103259.82+442202.2 (2567-54179-177), is reported here for the first time. SDSS J141118.31+481257.6 (1671-53446-0010) is identified as a galaxy at redshift 0.135 by the SDSS pipeline. However, it is actually an AM Canum Venaticorum star, which is a rare type of cataclysmic variable star, where a white dwarf accretes hydrogen-poor matter from a compact companion star \citep{anderson05}. We list these in table \ref{table:stars}. Finally, 10 objects suffer from various problems such as bad sky subtraction, and wrong flux calibration (i.e., flux level from the blue and red arms of the spectrograph do not match; table \ref{table:instrumental}). 

\section{Conclusions and Discussion}\label{s:conc}

We have introduced an outlier detection algorithm that is based on an unsupervised implementation of RF. By construction, the algorithm learns the most important features of the data and their interconnections; it is completely general and can be applied to imaging data, time-series, and other spectroscopic objects as well. Out of 2\,355\,926 galaxies which compose the input sample, we chose 400 galaxies with the highest weirdness score. We find objects with unusual emission line ratios, and complex velocity structures, extremely red objects, objects with extremely strong absorption lines (i.e. sodium and $\mathrm{H\delta}$), galaxies which host supernovae, or have rare emission lines. 

The algorithm finds a large variety of outliers arising from a large variety of physical environments. It is sensitive to the ratio between emission lines, thus to the radiation field that gas in a galaxy is exposed to, and to the metallically of the gas. It is sensitive to the widths of emission lines i.e., objects with extreme velocity dispersion, or with a number of different velocity components. It is sensitive to the shape of the continuum and to its connection to the strength of the emission lines and absorption lines. It recovers a number of rare phenomena that are subject to great attention in the astronomical community: $\mathrm{H\delta}$-strong galaxies, outflows and shocks, supernovae, galaxy-lenses, double-peaked emission lines which may be due to galaxy mergers or binary BHs. Only about a fifth of the objects we find were discovered in previous studies, using specific algorithms that isolate a single type of objects they were looking for (or searching by eye). Our algorithm detects all of them in one fell swoop. Despite how thoroughly SDSS has been explored before, we found two rare classes of galaxies that were never noticed or studied before, E+A galaxies with broad emission lines, and strong-$\mathrm{[NI]}$ emitting galaxies. 

Apart from these, the algorithm detects a number of different `instrumental' outliers, and can be used to detect pipeline failures in new or ongoing surveys. Most importantly, since the sole input to the algorithm is the data, with no physical rules or user specifications, and we find that it is able to detect some of the most interesting phenomena in the SDSS sample, this algorithm can be used to search for new objects that represent new phenomena -- searching for the truly unknown unknowns.

From a performance perspective, the increasing volume and complexity of astronomical data requires the development of new algorithms for detection, classification, and analysis of large datasets. The The Large Synoptic Survey Telescope (LSST) for example, is expected to conduct 32 trillion observations of 40 billion objects \citep{ivezic08}. These datasets can no longer be placed on a single compute node nor can they be analysed at once with a single processor. The framework of DFS has evolved as a possible solution to this challenge and parallelisation had become an essential methodology in it. The RF-based outlier detection we propose is completely parallel and operates naturally in the DFS framework. Furthermore, the algorithm is suitable for ongoing surveys where static data is replaced by data streams -- new data is chunked and decision trees are trained without the need to re-run the process on the entire dataset. RF's ability to operate with missing values is an additional important feature. Finally, the algorithm can work with an extremely large number of features with little to no cost in execution time.

\section*{Acknowledgments}
We thank our referee, Y. Ascasibar, for useful comments and suggestions that helped improve this manuscript.
We thank G. Ferland, N. Foerster Schreiber, A. Gal-Yam, O. Graur, D. Maoz, H. Netzer, P. Nugent, E. Ofek, J. X. Prochaska, and D. Watson who helped us identify some of the unusual objects we found, and gave us valuable advice regarding this manuscript. DP is partially supported by Grant No 2014413 from the United States-Israel Binational Science Foundation (BSF). 

The bulk of our computations was performed on the resources of the National Energy Research Scientific Computing Center, which is supported by the Office of Science of the U.S. Department of Energy under Contract No. DE-AC02-05CH11231. The  spectroscopic analysis was made using IPython \citep{perez07}. We also used the following Python packages:  pyspeckit\footnote{www.pyspeckit.bitbucket.org}, scikit-learn\footnote{www.scikit-learn.org} \citep{scikitlearn},
and astropy\footnote{www.astropy.org/}.

This work made extensive use of SDSS-III\footnote{www.sdss3.org} data. Funding for SDSS-III has been provided by the Alfred P. Sloan Foundation, the Participating Institutions, the National Science Foundation, and the U.S. Department of Energy Office of Science. SDSS-III is managed by the Astrophysical Research Consortium for the Participating Institutions of the SDSS-III Collaboration including the University of Arizona, the Brazilian Participation Group, Brookhaven National Laboratory, Carnegie Mellon University, University of Florida, the French Participation Group, the German Participation Group, Harvard University, the Instituto de Astrofisica de Canarias, the Michigan State/Notre Dame/JINA Participation Group, Johns Hopkins University, Lawrence Berkeley National Laboratory, Max Planck Institute for Astrophysics, Max Planck Institute for Extraterrestrial Physics, New Mexico State University, New York University, Ohio State University, Pennsylvania State University, University of Portsmouth, Princeton 
University, the Spanish Participation Group, University of Tokyo, University of Utah, Vanderbilt University, University of Virginia, University of Washington, and Yale University. 

\bibliographystyle{mn2e}
\bibliography{ref}

\clearpage

\appendix
\section{Outlying galaxies -- summary}\label{a:app}

We summarise all the galaxies we present in the paper in the next tables. We note that some galaxies are classified into multiple categories, in this case we present the galaxy in every class (and table) it belongs to.

\begin{table*}
\caption{Unusual velocity structure}\label{table:unus_vel}
\begin{tabular*}{0.99\linewidth}{@{\extracolsep{\fill}}l l l l l}
\hline
\hline
PLATE & MJD & FIBER & z & References \\
(1) & (2) & (3) & (4) & (5)\\
\hline
1323 & 52797 & 335 & 0.116 & \\
2117 & 54115 & 230 & 0.113 & \\
3788 & 55246 & 275 & 0.475 & \\
4767 & 55946 & 889 & 0.499 & SC13\\ 
7163 & 56593 & 697 & 0.599 & \\
\hline
\multicolumn{5}{l}{(1) -- SDSS plate ID of the galaxy.} \\                                                                           
\multicolumn{5}{l}{(2) -- SDSS MJD of the observation.} \\
\multicolumn{5}{l}{(3) -- SDSS fiber ID of the galaxy.} \\
\multicolumn{5}{l}{(4) -- Redshift from SDSS pipeline, whenever available we use $z_{NOQSO}$ which is more reliable. }\\
\multicolumn{5}{l}{(5) -- Previous studies of the galaxy, if exist: SC13 -- \citet{schirmer13} }
\end{tabular*}
\end{table*}

\begin{table*}
\caption{Galaxies with additional velocity structure near $\mathrm{H\alpha}$}\label{table:halpha_bumps}
\begin{tabular*}{0.99\linewidth}{@{\extracolsep{\fill}}l l l l l}
\hline
\hline
PLATE & MJD & FIBER & z & Comments\\
(1) & (2) & (3) & (4) & (5) \\
\hline
663 & 52145 & 306 & 0.041 & \\
1222 & 52763 & 341 & 0.028 & \\
1330 & 52822 & 391 & 0.111 & \\
1377 & 53050 & 441 & 0.009 & \\
1453 & 53084 & 289 & 4.5 & $z=0.038$\\
1724 & 53859 & 116 & 0.037 & \\
2022 & 53827 & 286 & 0.023 & \\
2511 & 53882 & 29 & 0.172 & \\
2660 & 54504 & 446 & 0.016 & \\
3929 & 55335 & 835 & 0.184 & \\
6138 & 56598 & 213 & 0.248 & \\
\hline
\multicolumn{5}{l}{(1) -- SDSS plate ID of the galaxy.} \\                                                                           
\multicolumn{5}{l}{(2) -- SDSS MJD of the observation.} \\
\multicolumn{5}{l}{(3) -- SDSS fiber ID of the galaxy.} \\
\multicolumn{5}{l}{(4) -- Redshift from SDSS pipeline, whenever available we use $z_{NOQSO}$ which is more reliable. }\\
\multicolumn{5}{l}{(5) -- Additional comments, if the redshift given by the SDSS pipeline is wrong we note the correct redshift.} \\
\end{tabular*}
\end{table*}

\begin{table*}
\caption{Double-peaked emission-line galaxies}\label{table:double_peaked}
\begin{tabular*}{0.99\linewidth}{@{\extracolsep{\fill}}l l l l l l}
\hline
\hline
PLATE & MJD & FIBER & z & Comments & References\\
(1) & (2) & (3) & (4) & (5) & (6)\\
\hline
397 & 51794 & 109 & 0.018 &  &  \\
1360 & 53033 & 186 & 0.19 &  & GE12, PI12 \\
1566 & 53003 & 429 & 0.246 & &  \\
2266 & 53679 & 422 & 0.016 & &  \\
4391 & 55866 & 835 & 0.069 & &  \\
4755 & 55660 & 45 & 0.433 & &  \\
5049 & 56103 & 273 & 0.229 & $z=0.608$ &  \\
5867 & 56034 & 304 & 0.15 & & XU09, SM10 \\
6788 & 56428 & 257 & 0.214 & $z=0.590$ &  \\
7047 & 56572 & 461 & 0.534 & &  \\
\hline
662	& 52147	& 180 & 5.7 & broad components, $z=0.259$ & ST03, WU04, ST08\\ 
733	& 52207 & 222 & 0.239 & broad components & ST03, WU04, BI07\\
838	& 52378	& 460 & 5.1 & broad components, $z=0.131$ & ST03, WU04, BI07\\
\hline
\multicolumn{6}{l}{(1) -- SDSS plate ID of the galaxy.} \\                                                                           
\multicolumn{6}{l}{(2) -- SDSS MJD of the observation.} \\
\multicolumn{6}{l}{(3) -- SDSS fiber ID of the galaxy.} \\
\multicolumn{6}{l}{(4) -- Redshift from SDSS pipeline, whenever available we use $z_{NOQSO}$ which is more reliable. }\\
\multicolumn{6}{l}{(5) -- Additional comments, if the redshift given by the SDSS pipeline is wrong we note the correct redshift.} \\
\multicolumn{6}{l}{(6) -- Previous studies of the galaxy, if exist: ST03 -- \citet{strateva03}, WU04 -- \citet{wu04}, BI07 -- \citet{bian07}, } \\
\multicolumn{6}{l}{\,\,\,\, ST08 -- \citet{strateva08}, XU09 -- \citet{xu09}, SM10 -- \citet{smith10}, GE12 -- \citet{ge12}, } \\ 
\multicolumn{6}{l}{\,\,\,\,PI12 -- \citet{pilyugin12}.} \\
\end{tabular*}
\end{table*}

\begin{table*}
\caption{Broad $\mathrm{[OIII]}$ emission line}\label{table:broad_oiii}
\begin{tabular*}{0.99\linewidth}{@{\extracolsep{\fill}}l l l l l l}
\hline
\hline
PLATE & MJD & FIBER & z & Comments & FWHM\\
(1) & (2) & (3) & (4) & (5) & (6)\\
\hline
2089 & 53498 & 416 & 0.081 & & 930 \\
2357 & 53793 & 115 & 0.109 & & 830 \\
3859 & 55246 & 651 & 0.177 & $z=0.54317$ & 620 \\
4295 & 55858 & 929 & 0.198 & & 710 \\
4776 & 55652 & 25 & 0.217 & $z=0.59778$ & 790 \\
5129 & 55864 & 169 & 0.185 & & 1000 \\
6031 & 56091 & 7 & 0.246 & $z=0.63461$ & 930 \\
6402 & 56334 & 838 & 0.153 & $z=0.5125$ & 1030 \\
6465 & 56279 & 757 & 0.08 & & 950 \\
6592 & 56535 & 653 & 0.18 & & 690 \\
6974 & 56442 & 627 & 0.506 & & 880 \\
7101 & 56659 & 293 & 0.088 & & 780 \\
6473 & 56363 & 672 & 0.116 & & 610 \\
\hline
\multicolumn{6}{l}{(1) -- SDSS plate ID of the galaxy.} \\                                                                           
\multicolumn{6}{l}{(2) -- SDSS MJD of the observation.} \\
\multicolumn{6}{l}{(3) -- SDSS fiber ID of the galaxy.} \\
\multicolumn{6}{l}{(4) -- Redshift from SDSS pipeline, whenever available we use $z_{NOQSO}$ which is more reliable. }\\
\multicolumn{6}{l}{(5) -- Additional comments, if the redshift given by the SDSS pipeline is wrong we note the correct redshift.} \\
\multicolumn{6}{l}{(6) -- Full width at half maximum of the $\mathrm{[OIII]}$ emission line in $\mathrm{km\,s^{-1}}$.} \\
\end{tabular*}
\end{table*}

\begin{table*}
\caption{$\mathrm{H\delta}$ strong galaxies}\label{table:ea_galaxies}
\begin{tabular*}{0.99\linewidth}{@{\extracolsep{\fill}}l l l l l l}
\hline
\hline
PLATE & MJD & FIBER & z & Comments & References\\
(1) & (2) & (3) & (4) & (5) & (6)\\
\hline
403 & 51871 & 219 & 0.156 & & GO03 \\
457 & 51901 & 166 & 0.116 & & GO03 \\
538 & 52029 & 413 & 0.264 & & GO03, GO04, BA05, GO07, ME13 \\
611 & 52055 & 128 & 0.130 & & GO03 \\
629 & 52051 & 561 & 0.154 & & GO03, BA05 \\
719 & 52203 & 555 & 0.157 & broad emission& \\
781 & 52373 & 607 & 0.167 & & GO07, ME13 \\
819 & 52409 & 369 & 0.136 & & GO07 \\
870 & 52325 & 536 & 0.143 & & GO07, ME13 \\
1185 & 52642 & 345 & 0.296 & & GO07, ME13 \\
1223 & 52781 & 386 & 0.004 & &  \\
1311 & 52765 & 359 & 0.163 & & GO07, ME13 \\
1366 & 53063 & 430 & 0.182 & & GO07, ME13 \\
1429 & 52990 & 416 & 0.17 & & GO07, ME13 \\
1463 & 53063 & 262 & 0.125 & broad emission &  \\
1574 & 53476 & 577 & 0.192 & & GO07, ME13 \\
1578 & 53496 & 1 & 0.173 & & GO07 \\
1583 & 52941 & 333 & 0.273 & & GO07, ME13 \\
1616 & 53169 & 135 & 0.008 & &  \\
1621 & 53383 & 336 & 0.185 & & ME13 \\
1658 & 53240 & 319 & 0.123 & & GO07, ME13 \\
1701 & 53142 & 496 & 0.28 & &  \\
1769 & 53502 & 17 & 0.228 & & ME13 \\
1989 & 53772 & 606 & 0.127 & & ME13 \\
2022 & 53827 & 88 & 0.168 & & ME13 \\
2123 & 53793 & 77 & 0.139 & & ME13 \\
2165 & 53917 & 68 & 0.296 & & ME13 \\
2168 & 53886 & 332 & 0.134 & & ME13 \\
2225 & 53729 & 345 & 0.178 & & ME13 \\
2292 & 53713 & 73 & 0.176 & & ME13 \\
2608 & 54474 & 8 & 0.177 & & ME13 \\
2789 & 54555 & 502 & 0.188 & & ME13 \\
2791 & 54556 & 321 & 0.185 & & ME13 \\
\hline
\multicolumn{6}{l}{(1) -- SDSS plate ID of the galaxy.} \\                                                                           
\multicolumn{6}{l}{(2) -- SDSS MJD of the observation.} \\
\multicolumn{6}{l}{(3) -- SDSS fiber ID of the galaxy.} \\
\multicolumn{6}{l}{(4) -- Redshift from SDSS pipeline, whenever available we use $z_{NOQSO}$ which is more reliable. }\\
\multicolumn{6}{l}{(5) -- Additional comments, if the redshift given by the SDSS pipeline is wrong we note the correct redshift.} \\
\multicolumn{6}{l}{(6) -- Previous studies of the galaxy, if exist: GO03 -- \citet{goto03}, GO04 -- \citet{goto04},} \\
\multicolumn{6}{l}{\,\,\,\,BA05 -- \citet{balogh05}, GO07 -- \citet{goto07}, ME13 -- \citet{melnick13}.}
\end{tabular*}
\end{table*}

\begin{table*}
\caption{Unusual emission lines}\label{table:cls}
\begin{tabular*}{0.99\linewidth}{@{\extracolsep{\fill}}l l l l l l}
\hline
\hline
PLATE & MJD & FIBER & z & Comments & References\\
(1) & (2) & (3) & (4) & (5) & (6)\\
\hline
2580 & 54092 & 470 & 0.101 & coronal lines & WA12 \\
4312 & 55511 & 732 & 0.059 & coronal lines &  \\
6473 & 56363 & 672 & 0.116 & coronal lines & \\
519	& 52283 & 415 & 0.023 & $\mathrm{[NI]}$ emission & \\
533 & 51994 & 537 & 0.025 & $\mathrm{[NI]}$ emission & \\
575 & 52319 & 200 & 0.29 & $\mathrm{[NI]}$ emission & \\
667 & 52163 & 205 & 0.013 & $\mathrm{[NI]}$ emission & \\
727 & 52207 & 504 & 0.043 & $\mathrm{[NI]}$ emission & \\
1324 & 53088 & 456 & 0.005 & $\mathrm{[NI]}$ emission & \\
1388 & 53119 & 128 & 0.362 & $\mathrm{[NI]}$ emission & \\
1816 & 53919 & 545 & 0.042 & $\mathrm{[NI]}$ emission & \\
2440 & 53818 & 135 & 0.136 & $\mathrm{[NI]}$ emission & \\
\hline
\multicolumn{6}{l}{(1) -- SDSS plate ID of the galaxy.} \\                                                                           
\multicolumn{6}{l}{(2) -- SDSS MJD of the observation.} \\
\multicolumn{6}{l}{(3) -- SDSS fiber ID of the galaxy.} \\
\multicolumn{6}{l}{(4) -- Redshift from SDSS pipeline, whenever available we use $z_{NOQSO}$ which is more reliable. }\\
\multicolumn{6}{l}{(5) -- Additional comments, if the redshift given by the SDSS pipeline is wrong due to broad emission lines we note the correct redshift.} \\
\multicolumn{6}{l}{(6) -- Previous studies of the galaxy, if exist: WA12 -- \citet{wang12}.} \\
\end{tabular*}
\end{table*}

\begin{table*}
\caption{Outliers on BPT diagram}\label{table:bpt}
\begin{tabular*}{0.99\linewidth}{@{\extracolsep{\fill}}llll|llll|llll}
\hline
\hline
PLATE & MJD & FIBER & z & PLATE & MJD & FIBER & z & PLATE & MJD & FIBER & z \\
(1) & (2) & (3) & (4) & (1) & (2) & (3) & (4) & (1) & (2) & (3) & (4) \\
\hline
267 & 51608 & 384 & 0.006 & 1162 & 52668 & 406 & 0.127 & 2131 & 53819 & 447 & 0.174 \\
280 & 51612 & 192 & 0.099 & 1198 & 52669 & 230 & 0.128 & 2167 & 53889 & 71 & 0.011 \\
284 & 51943 & 170 & 0.005 & 1216 & 52709 & 619 & 0.016 & 2218 & 53816 & 133 & 0.141 \\
291 & 51928 & 560 & 0.089 & 1291 & 52738 & 523 & 0.018 & 2228 & 53818 & 334 & 0.025 \\
330 & 52370 & 471 & 0.004 & 1311 & 52765 & 262 & 0.051 & 2233 & 53845 & 371 & 0.009 \\
336 & 51999 & 198 & 0.011 & 1311 & 52765 & 519 & 0.174 & 2238 & 54205 & 222 & 0.004 \\
444 & 51883 & 199 & 0.017 & 1324 & 53088 & 183 & 0.001 & 2245 & 54208 & 467 & 0.047 \\
446 & 51899 & 529 & 0.047 & 1325 & 52762 & 634 & 0.152 & 2266 & 53679 & 422 & 0.016 \\
446 & 51899 & 597 & 0.113 & 1327 & 52781 & 18 & 0.03 & 2286 & 53700 & 472 & 0.032 \\
450 & 51908 & 409 & 0.047 & 1333 & 52782 & 585 & 0.119 & 2297 & 53738 & 349 & 0.295 \\
460 & 51924 & 616 & 0.034 & 1366 & 53063 & 83 & 0.019 & 2322 & 53727 & 365 & 0.026 \\
462 & 51909 & 444 & 0.085 & 1382 & 53115 & 193 & 0.014 & 2421 & 54153 & 494 & 0.033 \\
468 & 51912 & 233 & 0.051 & 1427 & 52996 & 221 & 0.004 & 2494 & 54174 & 361 & 0.005 \\
486 & 51910 & 231 & 0.039 & 1451 & 53117 & 116 & 0.187 & 2497 & 54154 & 246 & 0.021 \\
488 & 51914 & 555 & 0.122 & 1467 & 53115 & 579 & 0.067 & 2513 & 54141 & 309 & 0.027 \\
489 & 51930 & 50 & 0.006 & 1513 & 53741 & 637 & 0.042 & 2525 & 54569 & 606 & 0.035 \\
499 & 51988 & 84 & 0.021 & 1601 & 53115 & 547 & 0.004 & 2528 & 54571 & 470 & 0.065 \\
516 & 52017 & 119 & 0.007 & 1602 & 53117 & 380 & 0.022 & 2561 & 54597 & 345 & 0.09 \\
519 & 52283 & 124 & 0.026 & 1612 & 53149 & 73 & 0.021 & 2638 & 54095 & 614 & 0.067 \\
521 & 52326 & 300 & 0.006 & 1615 & 53166 & 120 & 0.004 & 2648 & 54485 & 584 & 0.115 \\
546 & 52205 & 419 & 0.047 & 1625 & 53140 & 626 & 0.139 & 2742 & 54233 & 385 & 0.089 \\
553 & 51999 & 342 & 0.03 & 1626 & 53472 & 190 & 0.014 & 2748 & 54234 & 456 & 0.162 \\
556 & 51991 & 11 & 0.005 & 1655 & 53523 & 526 & 0.014 & 2772 & 54529 & 566 & 0.116 \\
574 & 52355 & 356 & 0.122 & 1672 & 53460 & 489 & 0.026 & 2776 & 54554 & 219 & 0.186 \\
603 & 52056 & 192 & 0.144 & 1692 & 53473 & 6 & 0.091 & 2779 & 54540 & 261 & 0.032 \\
616 & 52442 & 364 & 0.002 & 1701 & 53142 & 180 & 0.237 & 2880 & 54509 & 230 & 0.018 \\
624 & 52377 & 361 & 0.002 & 1707 & 53885 & 587 & 0.277 & 2961 & 54550 & 550 & 0.176 \\
647 & 52553 & 235 & 0.059 & 1719 & 53876 & 196 & 0.082 & 3676 & 55186 & 109 & 0.132 \\
657 & 52177 & 458 & 0.019 & 1724 & 53859 & 318 & 0.011 & 3788 & 55246 & 275 & 0.475 \\
735 & 52519 & 461 & 0.116 & 1740 & 53050 & 265 & 0.031 & 3942 & 55338 & 939 & 0.086 \\
753 & 52233 & 94 & 0.014 & 1740 & 53050 & 295 & 0.12 & 3946 & 55659 & 189 & 0.146 \\
761 & 54524 & 477 & 0.095 & 1745 & 53061 & 196 & 0.01 & 4181 & 55685 & 230 & 0.103 \\
764 & 52238 & 266 & 0.054 & 1751 & 53377 & 288 & 0.06 & 4317 & 55480 & 655 & 0.417 \\
778 & 54525 & 278 & 0.004 & 1761 & 53376 & 293 & 0.021 & 4320 & 55894 & 922 & 0.124 \\
778 & 54525 & 497 & 0.04 & 1782 & 53299 & 412 & 0.1 & 4352 & 55533 & 227 & 0.133 \\
788 & 52338 & 311 & 0.01 & 1787 & 54465 & 443 & 0.005 & 5022 & 55827 & 133 & 0.409 \\
844 & 52378 & 445 & 0.018 & 1788 & 54468 & 239 & 0.016 & 5309 & 55929 & 249 & 0.079 \\
853 & 52374 & 422 & 0.058 & 1802 & 53885 & 602 & 0.168 & 5357 & 55956 & 673 & 0.181 \\
853 & 52374 & 577 & 0.08 & 1805 & 53875 & 413 & 0.15 & 5428 & 56029 & 869 & 0.15 \\
861 & 52318 & 489 & 0.02 & 1814 & 54555 & 395 & 0.049 & 5430 & 56000 & 211 & 0.399 \\
880 & 52367 & 588 & 0.049 & 1975 & 53734 & 467 & 0.002 & 5994 & 56101 & 342 & 0.352 \\
896 & 52592 & 34 & 0.054 & 2001 & 53493 & 146 & 0.001 & 6000 & 56102 & 85 & 0.107 \\
900 & 52637 & 534 & 0.192 & 2003 & 53442 & 450 & 0.04 & 6140 & 56189 & 595 & 0.073 \\
934 & 52672 & 483 & 0.187 & 2016 & 53799 & 185 & 0.004 & 6293 & 56561 & 573 & 0.106 \\
939 & 52636 & 403 & 0.128 & 2016 & 53799 & 430 & 0.016 & 6402 & 56334 & 33 & 0.056 \\
939 & 52636 & 637 & 0.145 & 2016 & 53799 & 590 & 0.045 & 6415 & 56310 & 569 & 0.409 \\
952 & 52409 & 247 & 0.01 & 2019 & 53430 & 327 & 0.1 & 6431 & 56311 & 161 & 0.064 \\
955 & 52409 & 608 & 0.002 & 2025 & 53431 & 7 & 0.006 & 6474 & 56362 & 767 & 0.208 \\
964 & 52646 & 570 & 0.047 & 2026 & 53711 & 425 & 0.002 & 6514 & 56487 & 671 & 0.128 \\
1018 & 52672 & 359 & 0.004 & 2095 & 53474 & 347 & 0.01 & 6596 & 56331 & 812 & 0.101 \\
1039 & 52707 & 288 & 0.023 & 2097 & 53491 & 183 & 0.031 & 7050 & 56573 & 475 & 0.039 \\
1048 & 52736 & 424 & 0.092 & 2117 & 54115 & 351 & 0.008 & 7164 & 56597 & 405 & 0.466 \\
\hline
\multicolumn{12}{l}{(1) -- SDSS plate ID of the galaxy.} \\                                                                           
\multicolumn{12}{l}{(2) -- SDSS MJD of the observation.} \\
\multicolumn{12}{l}{(3) -- SDSS fiber ID of the galaxy.} \\
\multicolumn{12}{l}{(4) -- Redshift from SDSS pipeline, whenever available we use $z_{NOQSO}$ which is more reliable. }\\
\end{tabular*}
\end{table*}

\begin{table*}
\caption{Weak $\mathrm{H\alpha}$ emission}\label{table:snh}
\begin{tabular*}{0.99\linewidth}{@{\extracolsep{\fill}}l l l l l l}
\hline
\hline
PLATE & MJD & FIBER & z & $\mathrm{log\,[SII]/H\alpha}$  & $\mathrm{log\,[NII]/H\alpha}$\\
(1) & (2) & (3) & (4) & (5) & (6)\\
\hline
298 & 51955 & 204 & 0.012 & -0.026 & 0.062 \\
405 & 51816 & 603 & 0.021 & 0.007 & 0.161 \\
422	& 51878	& 80 & 0.094 & 0.117 & 0.027 \\
447 & 51877 & 376 & 0.059 & -0.181 & 0.218 \\
468 & 51912 & 523 & 0.035 & -0.022 & 0.156 \\
519	& 52283	& 415 & 0.023 & 0.200 & 0.230 \\
533 & 51994 & 537 & 0.025 & 0.088 & 0.055 \\
667	& 52163 & 205 & 0.013 & -0.014 & 0.021 \\
727 & 52207 & 504 & 0.043 & 0.047 & -0.141 \\
1251 & 52964 & 559 & 0.059 & 0.120 & 0.190 \\
1324 & 53088 & 456 & 0.005 & 0.130 & 0.214 \\
1327 & 52781 & 215 & 0.007 & 0.002 & -0.142 \\
1404 & 52825 & 251 & 0.048 & 0.019 & 0.007 \\
1430 & 53002 & 356 & 0.023 & 0.183 & 0.286 \\
1460 & 53138 & 244 & 0.028 & 0.076 & 0.220 \\
1707 & 53885 & 113 & 0.12 & 0.060 & 0.050 \\
1719 & 53876 & 89 & 0.034 & -0.084 & -0.170 \\
1816 & 53919 & 545 & 0.042 & 0.092 & 0.223 \\
2028 & 53818 & 60 & 0.014 & 0.129 & 0.125 \\
2105 & 53472 & 230 & 0.022 & 0.073 & 0.164 \\
2199 & 53556 & 78 & 0.04 & 0.090 & 0.101 \\
2440 & 53818 & 135 & 0.136 & 0.108 & 0.127 \\
2487 & 53852 & 117 & 0.047 & 0.020 & 0.236 \\
4745 & 55892 & 18 & 0.102 & 0.120 & 0.311 \\
5994 & 56101 & 342 & 0.352 & 0.054 & 0.297 \\
6138 & 56598 & 270 & 0.198 & 0.044 & 0.038 \\
6258 & 56238 & 282 & 0.183 & 0.018 & 0.009 \\
6261 & 56219 & 680 & 0.191 & 0.137 & 0.154 \\
7054 & 56575 & 500 & 0.138 & 0.171 & 0.063 \\
7057 & 56593 & 115 & 0.068 & 0.053 & -0.079 \\
\hline
\multicolumn{6}{l}{(1) -- SDSS plate ID of the galaxy.} \\                                                                           
\multicolumn{6}{l}{(2) -- SDSS MJD of the observation.} \\
\multicolumn{6}{l}{(3) -- SDSS fiber ID of the galaxy.} \\
\multicolumn{6}{l}{(4) -- Redshift from SDSS pipeline, whenever available we use $z_{NOQSO}$ which is more reliable. }\\
\multicolumn{6}{l}{(5) -- $\mathrm{log\,[SII]/H\alpha}$ measurement for the galaxy.}\\
\multicolumn{6}{l}{(6) -- $\mathrm{log\,[NII]/H\alpha}$ measurement for the galaxy.}
 \\
\end{tabular*}
\end{table*}

\begin{table*}
\caption{Sodium excess galaxies}\label{table:sod}
\begin{tabular*}{0.99\linewidth}{@{\extracolsep{\fill}}l l l l l}
\hline
\hline
PLATE & MJD & FIBER & z & References \\
(1) & (2) & (3) & (4) & (5)\\
\hline
267 & 51608 & 508 & 0.013 & \\
411 & 51817 & 362 & 0.072 & JE13 \\
412 & 52258 & 180 & 0.127 & \\
412 & 52258 & 492 & 0.047 & JE13 \\
427 & 51900 & 71 &  & \\
440 & 51885 & 227 & 0.117 & \\
468 & 51912 & 523 & 0.035 & \\
473 & 51929 & 492 & 0.085 & GO03, GO07, ME13 \\
512 & 51992 & 536 & 0.109 & \\
588 & 52045 & 622 & 0.044 & JE13 \\
591 & 52022 & 467 & 0.037 & \\
613 & 52345 & 393 & 0.008 & \\
667 & 52163 & 205 & 0.013 & JE13 \\
676 & 52178 & 420 & 0.016 & \\
731 & 52460 & 381 & 0.029 & JE13 \\
862 & 52325 & 108 & 0.103 &  \\
1055 & 52761 & 515 & 0.103 &  \\
1251 & 52964 & 559 & 0.059 & \\
1303 & 53050 & 443 & 0.011 & JE13 \\
1617 & 53112 & 454 & 0.107 &  \\
1620 & 53137 & 443 & 0.022 &  \\
1624 & 53386 & 594 & 0.009 &  \\
1672 & 53460 & 489 & 0.026 & JE13 \\
1707 & 53885 & 113 & 0.12 &  \\
1719 & 53876 & 89 & 0.034 & \\
1946 & 53432 & 350 & 0.005 &  \\
2199 & 53556 & 78 & 0.04 & JE13 \\
2120 & 53852 & 540 & 0.027 & JE13\\
2201 & 53904 & 263 & 0.114 &  \\
2235 & 53847 & 86 & 0.024 &  \\
2418 & 53794 & 116 & 0.035 & \\
2508 & 53875 & 389 & 0.012 & JE13 \\
2754 & 54240 & 372 & 0.013 & JE13 \\
2777 & 54554 & 201 & 0.037 & JE13 \\
2787 & 54552 & 606 & 0.044 & JE13 \\
7140 & 56569 & 428 & 0.065 &  \\
7161 & 56625 & 262 & 0.075 & \\
7161 & 56625 & 702 & 0.041 & \\
\hline
\multicolumn{5}{l}{(1) -- SDSS plate ID of the galaxy.} \\                                                                           
\multicolumn{5}{l}{(2) -- SDSS MJD of the observation.} \\
\multicolumn{5}{l}{(3) -- SDSS fiber ID of the galaxy.} \\
\multicolumn{5}{l}{(4) -- Redshift from SDSS pipeline, whenever available we use $z_{NOQSO}$ which is more reliable. }\\
\multicolumn{5}{l}{(5) -- Previous studies of the galaxy, if exist: GO03 -- \citet{goto03}, GO07 -- \citet{goto07},} \\ 
\multicolumn{5}{l}{\,\,\,\,JE13 -- \citet{jeong13}, ME13 -- \citet{melnick13}.} \\
\end{tabular*}
\end{table*}

\begin{table*}
\caption{Extremely red galaxies}\label{table:red}
\begin{tabular*}{0.99\linewidth}{@{\extracolsep{\fill}}l l l l l l}
\hline
\hline
PLATE & MJD & FIBER & z & Best template & Additional reddening \\
(1) & (2) & (3) & (4) & (5) & (6)\\
\hline
591 & 52022 & 467 & 0.037 & H$\alpha$ detected & $\mathrm{E}_{B-V}=0.64$ mag \\
676 & 52178 & 420 & 0.016 & H$\alpha$ detected & $\mathrm{E}_{B-V}=0.70$ mag \\
1251 & 52964 & 559 & 0.059 & LINER & $\mathrm{E}_{B-V}=0.65$ mag \\
1254 & 52972 & 234 & 0.013 & H$\alpha$ detected & $\mathrm{E}_{B-V}=0.92$ mag \\
1257 & 52944 & 314 & 0.012 &  MW dust cloud, $\mathrm{A_{r}}=4.42$ \\
1617 & 53112 & 454 & 0.107 & H$\alpha$ detected & $\mathrm{E}_{B-V}=0.38$ mag \\
1707 & 53885 & 113 & 0.12 & AGN + H$\alpha$ &  $\mathrm{E}_{B-V}=0.30$ mag \\
1719 & 53876 & 89 & 0.034 & AGN + H$\alpha$ & $\mathrm{E}_{B-V}=0.97$ mag \\
1946 & 53432 & 350 & 0.005 & H$\alpha$ detected & $\mathrm{E}_{B-V}=0.64$ mag \\
2120 & 53852 & 540 & 0.027 & H$\alpha$ detected & $\mathrm{E}_{B-V}=0.62$ mag \\
2418 & 53794 & 116 & 0.035 & AGN + H$\alpha$ & $\mathrm{E}_{B-V}=0.50$ mag \\
2508 & 53875 & 389 & 0.012 & H$\alpha$ detected & $\mathrm{E}_{B-V}=0.84$ mag \\
7035 & 56568 & 452 & 0.1 & H$\alpha$ detected & $\mathrm{E}_{B-V}=0.68$ mag\\
7048 & 56575 & 454 & 0.063 & H$\alpha$ detected & $\mathrm{E}_{B-V}=0.95$ mag\\
7056 & 56577 & 500 & 0.087 & AGN + H$\alpha$ & $\mathrm{E}_{B-V}=0.72$ mag\\
7152 & 56660 & 174 & 0.122 & Star-forming & $\mathrm{E}_{B-V}=0.70$ mag\\
7054 & 56575 & 500 & 0.138 & AGN + H$\alpha$ &  $\mathrm{E}_{B-V}=0.83$ mag\\
7057 & 56593 & 115 & 0.068 & LINER & $\mathrm{E}_{B-V}=0.63$ mag\\
7161 & 56625 & 262 & 0.075 & Star-forming & $\mathrm{E}_{B-V}=0.78$ mag \\
\hline
\multicolumn{6}{l}{(1) -- SDSS plate ID of the galaxy.} \\                                                                           
\multicolumn{6}{l}{(2) -- SDSS MJD of the observation.} \\
\multicolumn{6}{l}{(3) -- SDSS fiber ID of the galaxy.} \\
\multicolumn{6}{l}{(4) -- Redshift from SDSS pipeline, whenever available we use $z_{NOQSO}$ which is more reliable. } \\
\multicolumn{6}{l}{(5) -- Best fitting template from the SDSS galaxy composites by \citet{dobos12}.} \\ 
\multicolumn{6}{l}{(6) -- Additional dust reddening to be applied in order to obtain the continuum shape of the galaxy.} \\ 
\end{tabular*}
\end{table*}

\begin{table*}
\caption{Galaxies hosting supernovae}\label{table:sne}
\begin{tabular*}{0.99\linewidth}{@{\extracolsep{\fill}}l l l l l l}
\hline
\hline
PLATE & MJD & FIBER & z & References & Comments \\
(1) & (2) & (3) & (4) & (5) & (6)\\
\hline
424 & 51893 & 355 & 0.054 & GR15 & Ia\\
438 & 51884 & 462 & 0.117 & GR15 & Ia\\
472 & 51955 & 247 & 0.07 & MA03 & Ia \\
475 & 51965 & 626 & 0.031 & GR15 & Ia\\
966 & 52642 & 221 & 0.074 & GR15 & Ia\\
1304 & 52993 & 552 & 0.094 & GR15 & Ia \\
1392 & 52822 & 147 & 0.05 & GR15 & Ia\\
1462 & 53112 & 638 & 0.028 & GR15 & Ia\\
1744 & 53055 & 210 & 0.06 & GR15 & Ia\\
2199 & 53556 & 232 & 0.045 & GR15 & Ia\\
2202 & 53566 & 403 & 0.084 & GR15 & Ia\\
2222 & 53799 & 480 & 0.076 & GR15 & Ia\\
2430 & 53815 & 267 & 0.051 & GR15 & Ia\\
2744 & 54272 & 561 & 0.014 & GR15 & Ia\\
2792 & 54556 & 210 & 0.036 & GR15 & Ia\\
\hline
2281 & 53711 & 149 & 4.9 & & $z=0.094$, Ic +4 days \\
2281 & 53711 & 156 & 4.9 & & $z=0.211$, Ia, +3.2 days \\
4745 & 55892 & 18 & 0.102 &  & Ia +2 days \\
\hline
\multicolumn{6}{l}{(1) -- SDSS plate ID of the galaxy.} \\                                                                           
\multicolumn{6}{l}{(2) -- SDSS MJD of the observation.} \\
\multicolumn{6}{l}{(3) -- SDSS fiber ID of the galaxy.} \\
\multicolumn{6}{l}{(4) -- Redshift from SDSS pipeline, whenever available we use $z_{NOQSO}$ which is more reliable. } \\
\multicolumn{6}{l}{(5) -- Previous studies of the galaxy, if exist: MA03 -- \citet{madgwick03a}, GR15 -- \citet{graur15}.} \\ 
\multicolumn{6}{l}{(6) -- Additional comments: supernova type, correct redshift (if necessary), supernova age (if reported for the first time).} \\ 
\end{tabular*}
\end{table*}

\begin{table*}
\caption{Chance alignment -- galaxy and a nearby star}\label{table:chance_al}
\begin{tabular*}{0.99\linewidth}{@{\extracolsep{\fill}}l l l l}
\hline
\hline
PLATE & MJD & FIBER & z \\
(1) & (2) & (3) & (4) \\
269 & 51910 & 466 & 0.178 \\
691	& 52199 & 123 & 0.014 \\
967 & 52636 & 302 & 0.001 \\
1010 & 52649 & 301 & 0.022 \\
1784 & 54425 & 75 & 0.085 \\
2100 & 53713 & 106 & 0.024 \\
2477 & 54058 & 460 & 0.044 \\
2516 & 54241 & 391 & 0.035 \\
4350 & 55556 & 266 & 0.205 \\
5068 & 55749 & 474 & 0.161 \\
6138 & 56598 & 559 & 0.189 \\
6485 & 56342 & 655 & 0.023 \\
\hline
\multicolumn{4}{l}{(1) -- SDSS plate ID of the galaxy.} \\                                                                           
\multicolumn{4}{l}{(2) -- SDSS MJD of the observation.} \\
\multicolumn{4}{l}{(3) -- SDSS fiber ID of the galaxy.} \\
\multicolumn{4}{l}{(4) -- Redshift from SDSS pipeline, whenever available we use $z_{NOQSO}$ which is more reliable. }\\
\end{tabular*}
\end{table*}

\begin{table*}
\caption{Galaxy-galaxy gravitational lenses}\label{table:g_lens}
\begin{tabular*}{0.99\linewidth}{@{\extracolsep{\fill}}l l l l l}
\hline
\hline
PLATE & MJD & FIBER & z & References \\
(1) & (2) & (3) & (4) & (5) \\
\hline
328 & 52282 & 535 & 0.106 & BO08 \\
393 & 51794 & 456 & 0.1196 & BO08 \\
499 & 51988 & 5 & 0.123 & BO08 \\
615 & 52347 & 594 & 0.1428 & BO08 \\
807 & 52295 & 614 & 0.2803 &  BO08\\
969 & 52442 & 134 & 0.215 & BO08 \\
\hline
\multicolumn{5}{l}{(1) -- SDSS plate ID of the galaxy.} \\                                                                           
\multicolumn{5}{l}{(2) -- SDSS MJD of the observation.} \\
\multicolumn{5}{l}{(3) -- SDSS fiber ID of the galaxy.} \\
\multicolumn{5}{l}{(4) -- Redshift from SDSS pipeline, whenever available we use $z_{NOQSO}$ which is more reliable. }\\
\multicolumn{5}{l}{(5) -- Previous studies of the galaxy, if exist: BO08 -- \citet{bolton08}.} \\
\end{tabular*}
\end{table*}

\begin{table*}
\caption{Multiple emission-line systems}\label{table:multi_lines}
\begin{tabular*}{0.99\linewidth}{@{\extracolsep{\fill}}l l l l l}
\hline
\hline
PLATE & MJD & FIBER & z & $\mathrm{z_{sec}}$ \\
(1) & (2) & (3) & (4) & (5) \\
345 & 51690 & 229 & 0.113 & 0.1230 \\
746 & 52238 & 531 & 0.103 & 0.0463 \\
1420 & 53146 & 528 & 0.117 & 0.0627 \\
1421 & 53149 & 253 & 0.133 & 0 \\
1543 & 53738 & 574 & 0.125 & 0.0944 \\
1699 & 53148 & 495 & 0.148 & 0.1069 \\
1762 & 53415 & 435 & 0.13 & 0.0019 \\
4079 & 55363 & 773 & 0.133 & 0.733 \\
5399 & 55956 & 245 & 0.276 & 0.0425 \\
\hline
\multicolumn{5}{l}{(1) -- SDSS plate ID of the galaxy.} \\                                                                           
\multicolumn{5}{l}{(2) -- SDSS MJD of the observation.} \\
\multicolumn{5}{l}{(3) -- SDSS fiber ID of the galaxy.} \\
\multicolumn{5}{l}{(4) -- Redshift from SDSS pipeline, whenever available we use $z_{NOQSO}$ which is more reliable. }\\
\multicolumn{5}{l}{(5) -- Redshift of the second set of emission lines.} \\
\end{tabular*}
\end{table*}

\begin{table*}
\caption{Stars identified as galaxies}\label{table:stars}
\begin{tabular*}{0.99\linewidth}{@{\extracolsep{\fill}}l l l l l}
\hline
\hline
PLATE & MJD & FIBER & Comments & References \\
(1) & (2) & (3) & (4) & (5)\\
465 & 51910 & 482 & & \\
550 & 51959 &	433 & & \\
1237 &	52762 &	621 & & \\
1246 &	54478 &	520 & & \\
1247 &	52677 &	229 & & \\ 
1247 &	52677 &	229 & & \\
1247 &	52677 &	259 & & \\
1250 &	52930 &	543 & & \\
1251 &	52964 &	193 & & \\ 
1256 &	52902 &	585 & & \\
1256 &	52902 &	402 & & \\
1256 &	52902 &	506 & & \\ 
1256 &	52902 &	362 & & \\
1256 &	52902 &	38 & & \\ 
1256 &	52902 &	636 & & \\
1257 &	52944 &	253 & & \\ 
1257 &	52944 &	300 & & \\ 
1257 &	52944 &	209 & & \\
1257 &	52944 &	244 & & \\ 
1642 &	53115 &	81 & & \\
1999 &	53503 &	224 & & \\
2266 & 53679 & 82 & & \\ 
2445 &	54573 &	16 & & \\ 
2538 &	54271 &	381 & & \\
2538 &	54271 &	261 & & \\
2538 &	54271 &	359 & & \\
2538 &	54271 &	59 & & \\
2546 &	54625 &	124 & & \\
2555 &	54265 &	8 & & \\
2555 &	54265 &	10 & & \\
2568 &	54153 &	3 & & \\
2714 &	54208 &	222 & & \\
2799 &	54368 &	3 & & \\
2806 &	54425 &	5 & & \\
2807 &	54433 &	195 & & \\
2815 &	54414 &	228 & & \\ 
3175 &	54828 &	372 & & \\ 
3290 &	54941 &	314 & & \\
3324 &	54943 &	272 & & \\
3480 &	54999 &	228 & & \\
5042 &	55856 &	436 & & \\
550	& 51959 & 433 & WD & GI11 \\
1237 & 52762 & 621 & WD & KL13 \\
1679 & 53149 & 616 & WD & KL13 \\
1702 & 53144 & 178 & WD & KL13 \\
2567 & 54179 & 177 & WD & \\
1671 & 53446 & 10 & AM CVn & AN05 \\
\hline
\multicolumn{5}{l}{(1) -- SDSS plate ID of the galaxy.} \\                                                                           
\multicolumn{5}{l}{(2) -- SDSS MJD of the observation.} \\
\multicolumn{5}{l}{(3) -- SDSS fiber ID of the galaxy.} \\
\multicolumn{5}{l}{(4) -- Additional comments.} \\
\multicolumn{5}{l}{(5) -- Previous studies, if exist: AN05 -- \citet{anderson05}, GI11 -- \citet{girven11}, KL13 -- \citet{kleinman13}.} \\
\end{tabular*}
\end{table*}

\begin{table*}
\caption{Bad spectra}\label{table:instrumental}
\begin{tabular*}{0.99\linewidth}{@{\extracolsep{\fill}}l l l l l}
\hline
\hline
PLATE & MJD & FIBER & z & Comments \\
(1) & (2) & (3) & (4) & (5)\\
\hline
434 & 51885 & 639 & 0.073 & sky subtraction \\
446 & 51899 & 139 & 0.754 & redshift estimation \\
468 & 51912 & 233 & 0.051 & sky subtraction \\
473 & 51929 & 374 & 0.271 & flux calibration \\
767 & 52252 & 313 & 0.014 & flux calibration \\
1019 & 52707 & 261 & 0.008 & H$\alpha$ emission is cut off \\
1360 & 53003 & 633 & 0.124 & sky emission \\
1809 & 53792 & 99 & 0.058 & H$\alpha$ emission is cut off \\
2134 & 53876 & 45 & 0.053 & H$\alpha$ emission is cut off \\ 
5001 & 55719 & 40 & 0.201 & flux calibration \\
\hline
\multicolumn{5}{l}{(1) -- SDSS plate ID of the galaxy.} \\                                                                           
\multicolumn{5}{l}{(2) -- SDSS MJD of the observation.} \\
\multicolumn{5}{l}{(3) -- SDSS fiber ID of the galaxy.} \\
\multicolumn{4}{l}{(4) -- Redshift from SDSS pipeline, whenever available we use $z_{NOQSO}$ which is more reliable.} \\
\multicolumn{5}{l}{(5) -- Description of the source of contamination.} \\ 
\end{tabular*}
\end{table*}

\end{document}